\documentclass{aa}
\usepackage{textcomp}
\usepackage{graphicx}
\usepackage{amsmath}
\usepackage{amssymb}
\usepackage{bm}
\usepackage{upgreek}
\usepackage{IEEEtrantools}
\usepackage{multirow}
\usepackage[dvipsnames]{xcolor}
\usepackage[colorlinks=true,allcolors=blue,urlcolor=blue]{hyperref}
\usepackage{txfonts}
\usepackage[normalem]{ulem} 
\usepackage{aas_macros}

\newcommand{\sect}[1]{\text{Section~\ref{#1}}}
\newcommand{\fig}[1]{\text{Figure~\ref{#1}}}
\newcommand{\tab}[1]{\text{Table~\ref{#1}}}

\newcommand{\multitd}{\texttt{Multi3D}}
\newcommand{\multi}{\texttt{MULTI}}
\newcommand{\balder}{\texttt{Balder}}
\newcommand{\cobold}{\texttt{CO$^{5}$BOLD}}
\newcommand{\scate}{\texttt{Scate}}
\newcommand{\blue}{\texttt{Blue}}
\newcommand{\marcs}{\texttt{MARCS}}
\newcommand{\atlas}{\texttt{ATLAS9}}
\newcommand{\stagger}{\texttt{Stagger}}
\newcommand{\atmo}{\texttt{ATMO}}
\newcommand{\sme}{\texttt{SME}}
\newcommand{\pysme}{\texttt{PySME}}

\newcommand{\bsr}{\texttt{BSR}}
\newcommand{\grasp}{\texttt{GRASP}}
\newcommand{\civ}{\texttt{CIV3}}

\newcommand{\lgeps}[1]{A(\mathrm{#1})}
\newcommand{\lggf}{\log{gf}}
\newcommand{\nm}{\mathrm{nm}}
\newcommand{\ev}{\mathrm{eV}}
\newcommand{\dex}{\mathrm{dex}}
\newcommand{\kms}{\mathrm{km\,s^{-1}}}
\newcommand{\vsini}{\mathrm{V_{\mathrm{rot}}\sin{\iota}}}
\newcommand{\tcond}{T_{\mathrm{cond}}}
\newcommand{\kelvin}{\mathrm{K}}
\newcommand{\vmic}{\xi_{\mathrm{1D}}}

\begin{document} 

\title{The solar sulphur abundance in view of
large-scale atomic structure calculations
and 3D non-LTE models}
\author{A.~M.~Amarsi\inst{\ref{uu1}}
\and
W.~Li\inst{\ref{naoc}}
\and
N.~Grevesse\inst{\ref{liege1},\ref{liege2}}
\and
A.~J.~G.~Jurewicz\inst{\ref{asu},\ref{dartmouth}}}

\institute{\label{uu1}Theoretical Astrophysics, 
Department of Physics and Astronomy,
Uppsala University, Box 516, SE-751 20 Uppsala, Sweden
\and
\label{naoc}National Astronomical Observatories,
Chinese Academy of Sciences,
Beijing 100012, PR China
\and
\label{liege1}Centre Spatial de Li\`ege, Universit\'e de Li\'ege,
avenue Pr\'e Aily,
B-4031 Angleur-Li\`ege, Belgium
\and
\label{liege2}Space sciences, Technologies and Astrophysics Research (STAR)
Institute,
Universit\'e de Li\`ege, All\'ee du 6 ao\^ut, 17, B5C,
B-4000 Li\`ege, Belgium
\and
\label{asu}Busek Center for Meteorite Studies,
Arizona State University, Tempe, Arizona 85287-6004, USA
\and
\label{dartmouth}Department of Earth Sciences, Dartmouth College, Hanover, New
Hampshire 03755, USA}

\abstract{The solar chemical composition is a fundamental yardstick in
astrophysics and the topic of heated debate in recent literature.  We
re-evaluate the abundance of sulphur in the photosphere by studying seven
\ion{S}{I} lines in the solar disc-centre intensity spectrum.  Our analysis
considers independent sets of experimental and theoretical oscillator strengths
together with, for the first time, three-dimensional non-local thermodynamic
equilibrium (3D non-LTE) \ion{S}{I} spectrum synthesis.  Our best estimate is
$\lgeps{S}=7.06\pm0.04$, which is $0.06\,\dex$ to $0.10\,\dex$ lower than that
in commonly-used compilations of the solar chemical composition.  Our lower
solar sulphur abundance deviates from that in CI chondrites, and thereby
supports the case for a systematic difference
between the composition of the solar photosphere and of CI chondrites 
that is correlated with $50\%$ condensation temperature. We
suggest that precise laboratory measurements of \ion{S}{I} oscillator strengths
and abundance analyses using 3D magnetohydrodynamic 
models of the solar photosphere be
conducted to further substantiate our conclusions.}


\date{Received 3 August 2025 / Accepted 5 September 2025}
\maketitle

\section{Introduction}
\label{introduction}

The solar chemical composition is a fundamental yardstick in astrophysics,
against which other cosmic objects are most often compared
\citep[e.g.][]{1998SSRv...85..161G,2021A&A...653A.141A}.
In this context, there is great interest
for precise constraints on the solar sulphur abundance.
Sulphur is an $\alpha$-element
and thus interesting as a tracer of star formation
in different stellar populations in the Milky Way
\citep[e.g.][]{2021A&A...647A.162P,2022A&A...657A..29L,
2023A&A...671A.137L} and in other galaxies \citep[e.g.][]{2023MNRAS.521.1969D,
2024A&A...685A..81G,2025arXiv250614736P}.
Additionally, sulphur has diagnostic power for studying 
gas-dust fractionation processes in protoplanetary discs
around young stars \citep[e.g.][]{2019ApJ...885..114K,2024MNRAS.528..388K}
and in circumbinary discs around evolved stars
\citep[e.g.][]{2025MNRAS.538.1339M,mohorian_submitted}.
Sulphur is also relevant for studying the Sun itself.
Within the Sun, sulphur
is a non-negligible contributor to the opacity in the radiative zone
(see Figure 2 of \citealt{2025SoPh..300...97B}).
Thus sulphur may have a role in the long-debated
solar modelling problem \citep[e.g.][]{2021LRSP...18....2C}.
Sulphur may also be interesting 
for studies of the upper solar atmosphere.
As an element of intermediate first ionisation potential
(FIP; $E_{\mathrm{ion}}=10.36\,\ev$),
sulphur behaves like a high-FIP
element in the closed loop solar corona,
and like a low-FIP element in the solar wind,
and this can help shed light on the fractionation physics
\citep{2019ApJ...879..124L}.

In particular, recent analyses have put the spotlight on 
systematic differences
between the solar photospheric composition and the abundances inferred from CI
chondrites \citep[e.g.][]{2010MNRAS.407..314G,
2021A&A...653A.141A,2024M&PS...59.3193J}.  
Namely, either refractory elements
(with $50\%$ condensation temperature $\tcond\gtrsim1200\,\kelvin$), are
depleted in CI chondrites, or moderately-volatile elements
($400\lesssim\tcond\lesssim1200\,\kelvin$) are enhanced in CI chondrites,
compared to the photosphere, by up to around $20\%$. 
The measurements cannot constrain between these two scenarios
(enrichment versus depletion)
because that information is lost by converting the 
CI abundances to the solar scale, which in this work is
done by setting 
$\lgeps{Si}=7.51$\footnote{$\lgeps{X}\equiv\log_{10}N_{\mathrm{X}}/
N_{\mathrm{H}}+12$} from \citet{2021A&A...653A.141A}.
Nevertheless, these systematic differences call into
question the assumption that CI chondrites are a precise reflection of the
composition the protosolar nebula; and moreover the degree of depletion may help
constrain models of how the solar system formed \citep{2018ApJS..238...11D}.
Sulphur is a key element in this discussion
as it is a moderately-volatile element
($\tcond=672\,\kelvin$; \citealt{2019AmMin.104..844W}). There are only few
elements with such low $\tcond$ that have
precise photospheric abundance measurements;
and in the critical compilation of \citet{2021A&A...653A.141A},
sulphur has the lowest nominal error bar of just $0.03\,\dex$.
Thus, systematic errors
in the solar sulphur abundance could have a disproportionate
effect on the systematic differences between the solar
photospheric composition and CI chondrites, after taking 
the stipulated uncertainties into account.

The solar sulphur abundance from \citet{2021A&A...653A.141A},
$\lgeps{S}=7.12\pm0.03$,
comes from the analysis of \citet{2015A&A...573A..25S}.
This was based on spectrum synthesis of eight \ion{S}{I} lines
with the code \scate{} \citep{2011A&A...529A.158H}
using a three-dimensional (3D) radiation-hydrodynamics
simulation of the solar photosphere with the code \stagger{}
\citep{2018MNRAS.475.3369C,2024ApJ...970...24S}.
They compared synthetic equivalent widths
against measurements of solar disc-centre intensity
atlases \citep{1973apds.book.....D,1984SoPh...90..205N}
to obtain 3D LTE abundances.
To correct for departures from local thermodynamic
equilibrium (LTE), the authors calculated and applied
1D non-LTE versus 1D LTE abundance corrections 
using the model of \citet{2005PASJ...57..751T} to obtain
the final 3D LTE + 1D non-LTE result.

The purpose of this study is to revisit 
the solar sulphur abundance with consideration for
two possible sources of 
systematic error in the analysis of \citet{2015A&A...573A..25S}.
The first of these is the treatment of departures from LTE;
specifically, \citet{2015A&A...573A..25S}
carried out an inconsistent 3D LTE + 1D non-LTE
analysis,
and employed a now outdated non-LTE model atom 
employing the antiquated Drawin recipe for inelastic collisions
with neutral hydrogen
(see \citealt{2011A&A...530A..94B} for a discussion of this topic).
Here we carry out a consistent 3D non-LTE analysis,
based on a carefully constructed model atom
that uses modern atomic data.
The second possible source of systematic error
is the set of oscillator strengths for the diagnostic \ion{S}{I} lines;
five of their eight \ion{S}{I} lines do not have
precise laboratory measurements, and thus the
analysis of these lines is based on theoretical data.
Here we consider three independent sets of atomic structure calculations,
including new calculations from \citet{li_submitted},
to demonstrate that there are significant
systematic uncertainties in the atomic data.
Ultimately, we argue that the solar sulphur abundance
in \citet{2021A&A...653A.141A} is overestimated
and advocate $\lgeps{S}=7.06\pm0.04$.

\section{Method}
\label{method}

\begin{figure}
    \begin{center}
        \includegraphics[scale=0.325]{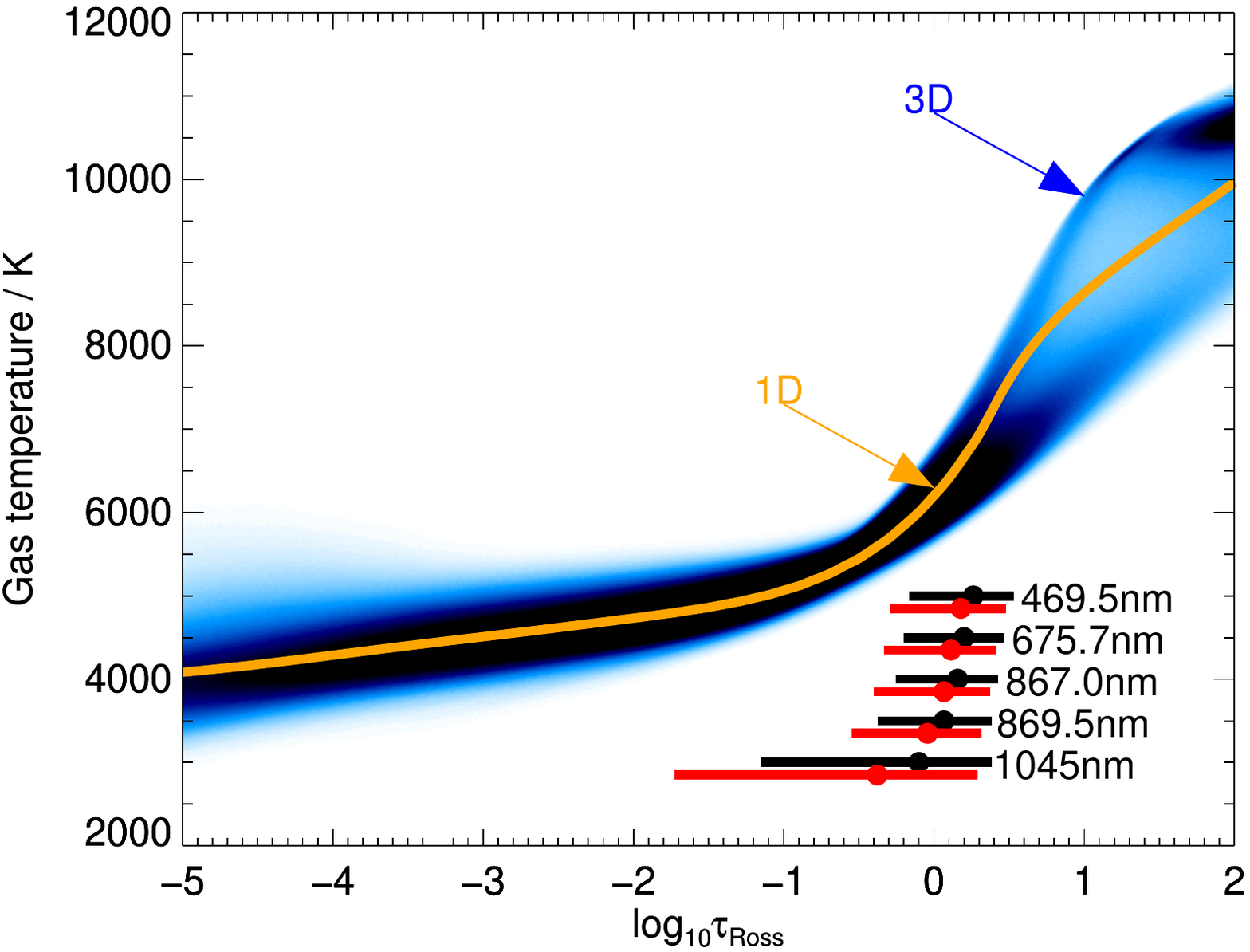}
        \caption{Gas temperature distribution
        with vertical logarithmic optical depth
        in the 3D model atmosphere (blue).
        The $T-\log\tau$ relation of the corresponding
        1D model atmosphere is also shown (orange).
        The full widths at half maximum of the 
        line-profile integrated 1D non-LTE contribution functions 
        to the line depressions
        \citep{2015MNRAS.452.1612A} are shown
        for the diagnostic \ion{S}{I} lines
        (the line-averaged result is shown for the 
        \ion{S}{I} $1045\,\nm$ triplet).
        These are presented for both the disc-centre intensity (black)
        and disc-integrated flux (red)
        with circles indicating the positions of the peaks.}
        \label{fig:atmos}
    \end{center}
\end{figure}

\begin{figure*}
    \begin{center}
        \includegraphics[scale=0.325]{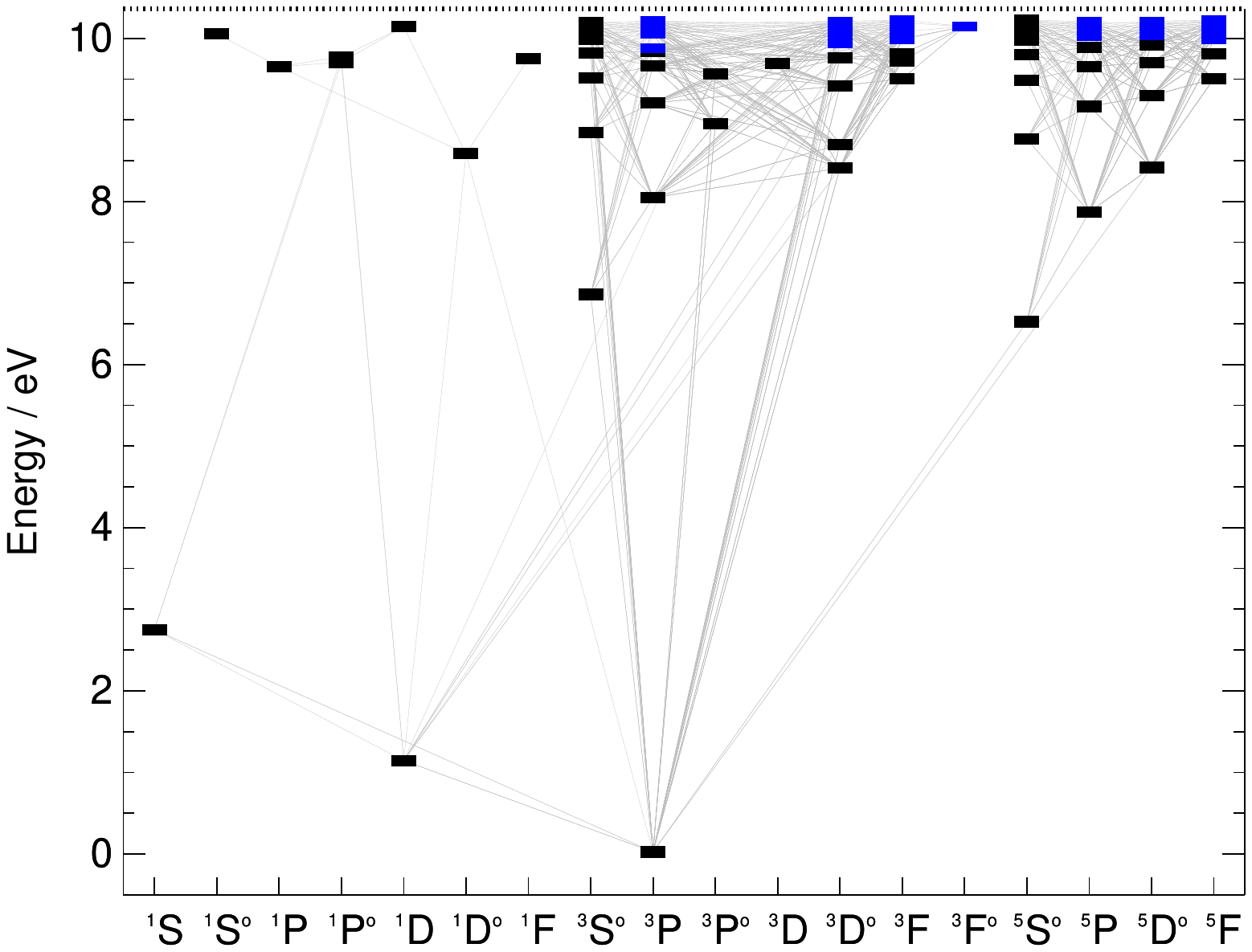}
        \includegraphics[scale=0.325]{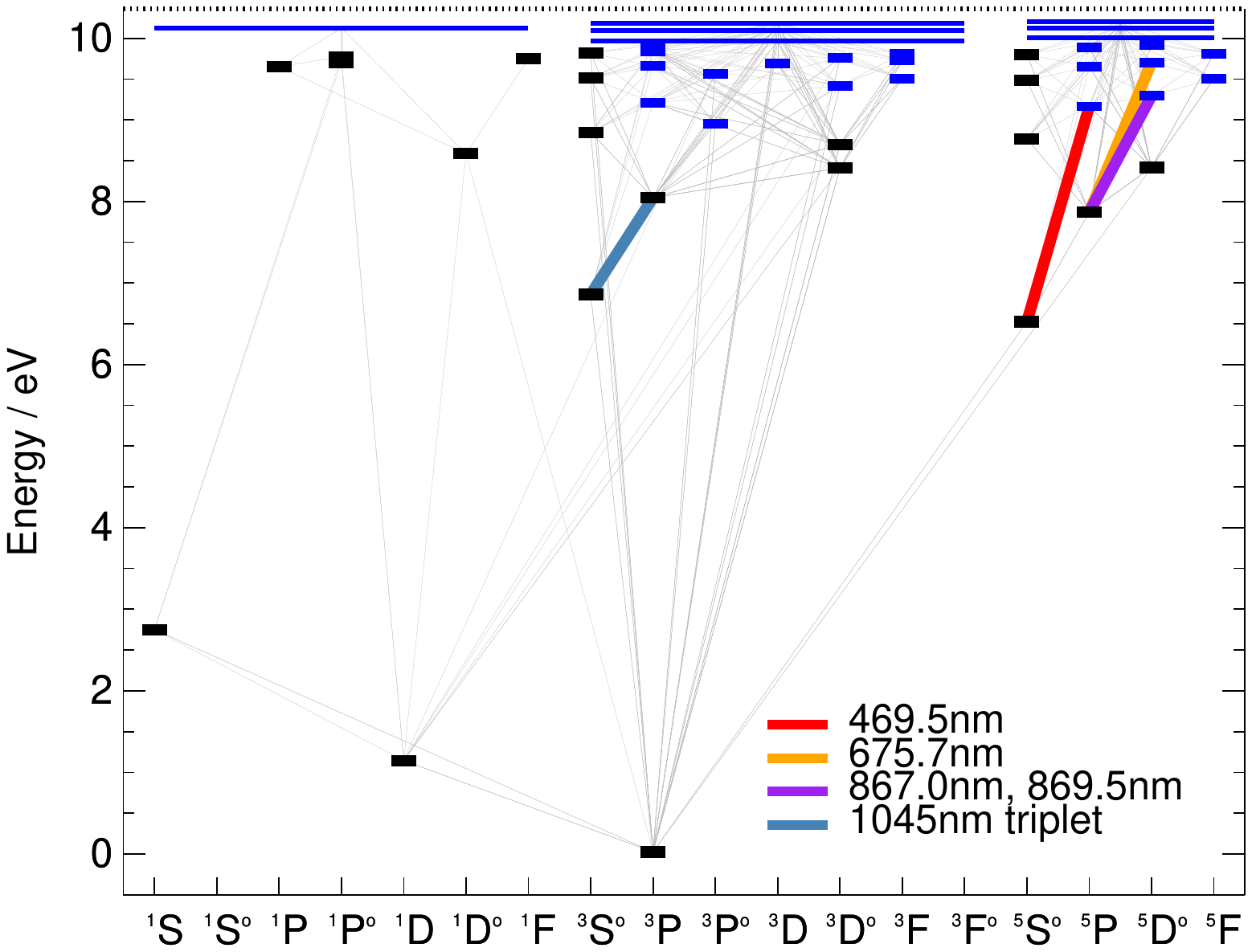}
        \caption{Grotrian diagram for the comprehensive (left) and
        reduced (right) model atoms used in this work.
        Terms for which fine-structure are unresolved
        are shown as short blue horizontal lines.
        Super levels in the reduced model atom are shown
        as long blue horizontal lines.
        Transitions shown as slanted grey lines;
        those used as abundance diagnostics are shown
        as coloured slanted lines
        in the right panel and labelled in the legend.}
        \label{fig:atom}
    \end{center}
\end{figure*}

\subsection{Model atmosphere and spectrum synthesis}
\label{method3n}

The synthetic spectra were calculated by post-processing
radiative transfer calculations of snapshots of
a 3D radiation-hydrodynamics simulation of the solar photosphere. 
This model atmosphere was calculated using the code \stagger{}
\citep{2018MNRAS.475.3369C,2024ApJ...970...24S}
with a resolution that
is comparable to that of the \stagger{}-grid
\citep{2013A&A...557A..26M},
but spanning around a day of solar time.
It is the same model that has been used 
in our continuing series of papers on the solar chemical composition 
beginning with \citet{2018A&A...616A..89A}
(see \citealt{2024A&A...690A.128A} and references therein).

The post-processing calculations were performed using the 3D LTE 
code \scate{} \citep{2011A&A...529A.158H},
and the 3D non-LTE code 
\balder{} \citep{2018A&A...615A.139A}; the latter
is a modified version of 
\multitd{} \citep{2009ASPC..415...87L_short}.
The model atom used by \balder{} is described in
\sect{methodatom} below.
For both sets of calculations, the model atmosphere
was downsampled in the two horizontal dimensions by a factor of $3^{2}$
and refined in the vertical for better sampling of the layers
with steep temperature gradient where the continuum photons escape
\citep[][]{2024A&A...688A.212R}.
The 3D non-LTE calculations using the \balder{} code
were carried out on eight snapshots of the 3D model atmosphere.
For each snapshot,
five different values of sulphur abundances
in steps of $0.2\,\dex$ were used
and the mean radiation field
was calculated using the $26$ ray angle quadrature
that is described in \citet{2024A&A...690A.128A}.
Background line opacities were pre-computed
using the code \blue{} \citep{2023A&A...677A..98Z}.
The 3D non-LTE to 3D LTE intensity ratios were computed and
applied to 3D LTE intensities calculated with \scate{}.
These latter calculations were performed
with finer temporal and abundance resolution,
namely $17$ snapshots, and $11$ sulphur abundances
using steps of $0.05\,\dex$.
The total and continuous spectra were averaged across these
multiple snapshot and then continuum-normalised.

To quantify the 3D effects (for example, the 
3D non-LTE versus 1D non-LTE abundance differences), 
post-processing calculations with the \scate{}
and \balder{} codes were also performed
on a 1D model atmosphere.
This model
was calculated with the same equation of state and treatment
of opacities as the 3D model,
but with convection now described via mixing length theory
(such \atmo{} models are described in the Appendix of 
\citealt{2013A&A...557A..26M}).
In the 1D case, the synthetic spectra were calculated assuming
a depth-independent microturbulence of $\vmic=1\,\kms$.
This fudge parameter accounts for broadening
caused by velocity gradients in the stellar atmosphere
on scales much smaller than one optical depth.
These gradients are not predicted by 1D models,
but they naturally emerge in the 3D simulations
\citep[e.g.][]{2009LRSP....6....2N}.
Thus, no microturbulence or other free
parameters (apart from the sulphur abundance)
were employed in the 3D LTE or 3D non-LTE spectrum syntheses.

We illustrate the temperature-$\tau$ relations of the 3D and 1D models
in \fig{fig:atmos}.
Comparisons with other 1D models as well as 
a horizontally- and temporally-averaged 3D model
can be found in Figure 2 of 
\citet{2021A&A...656A.113A}, for example.

\subsection{Model atom}
\label{methodatom}

We illustrate the levels and lines of the comprehensive
and reduced models in \fig{fig:atom},
and describe their construction below.
The same models were
recently used in \citet{2025MNRAS.538.1339M,mohorian_submitted} 
and \citet{2025arXiv250522615C};
an early version was used in
\citet{2024A&A...689A..36K}.

A comprehensive model atom was constructed
for neutral sulphur using fine structure levels
(namely, resolving L, S, and J; hereafter LSJ) up to 
$E=9.93\,\ev$ from \citet{1990JPCRD..19..821M}.
These data were extracted from the NIST Database
\citep{2020Atoms...8...56R} prior to
the recent update of \citet{2024ApJS..274...32C}.
The fine structure energies for the $\mathrm{5d\,^{5}D^{o}}$, 
$\mathrm{7p\,^{5}P}$, and $\mathrm{6d\,^{5}D^{o}}$ 
terms were updated using
the Kurucz database \citep{1995ASPC...78..205K}.
Higher LS terms were included from The Opacity Project 
\citep{1992RMxAA..23..107C,1995oppr.book.....S}, 
converting the reported quantum defects 
and experimental ionisation limits into energies relative to the ground state
in the way described in \citet{2017MNRAS.464..264A}.
Since the ionisation energy of sulphur is high
($10.36\,\ev$), neutral sulphur is the majority species
and so only the four lowest 
LS terms of singly ionised sulphur (taken from the NIST database
and collapsed) were included in the model.
Overall, the comprehensive model has $146$ LSJ levels or LS terms,
of which $142$ are of neutral sulphur.
Without any fine structure this would drop to $84$ LS terms,
comparable in complexity to the model presented in \citet{2024ARep...68.1159K}
that contains $64$ LS terms of neutral sulphur.

The bound-bound radiative transitions
were mainly taken from \citet{2009JPCRD..38..171P}
via the NIST database, most of which are based on
\citet{2006JPhB...39.2861Z}. 
These were supplemented with data from
The Opacity Project; where necessary,
these LS data were redistributed across the LSJ levels 
using the tables in \citet{1973asqu.book.....A}
to ensure a realistic description of the radiation
field and thus the radiative rates for bound-bound transitions.
With this combined data set in hand,
natural broadening parameters were calculated by summations
of the Einstein $A$ coefficients, and
Stark broadening parameters were extracted from the Kurucz
database. Hydrogen collision broadening parameters were determined
by interpolating the extended
tables\footnote{Available at \url{https://github.com/barklem/abo-cross}.} of
\citet{1995MNRAS.276..859A}, \citet{1997MNRAS.290..102B},
and \citet{1998MNRAS.296.1057B}; when the line parameters
were outside of these grids, 
the classical Uns\"{o}ld recipe was used instead, 
with enhancement factor of $2.0$.
The bound-free radiative transitions were taken from 
The Opacity Project, taking care to connect
the neutral sulphur levels to the correct ionic core,
shifting the provided wavelength basis such that
the threshold wavelengths were consistent with
the experimental energies from 
the NIST Database,
and (following \citealt{2008ApJ...682.1376B})
smoothing the sharp resonances in the 
data using a Gaussian filter.

For the collisions, the data for excitation by electron
impact were mainly taken from OpenADAS
\citep{2011AIPC.1344..179S}, file \texttt{ls\#s0.dat}.
For missing allowed transitions these were supplemented with calculations
based on the recipe of \citet{1962PPS....79.1105S}.
Missing forbidden transitions were crudely estimated
using the recipe of \citet{1962ApJ...136..906V}
taking $gf=10^{-3}$; our tests
found that these transitions had a negligible impact on
the statistical equilibrium.
The data for ionisation by electron impact were
estimated using the recipe of \citet{1973asqu.book.....A}.
Excitation by neutral hydrogen impact as well as charge transfer
involving different excited levels of neutral sulphur
($\mathrm{S+H \leftrightarrow S^{+}+H^{-}}$)
were taken from the asymptotic model calculations of
\citet{2020ApJ...893...59B}.  
As motivated by \citet[][]{2018A&A...616A..89A}
and subsequent studies,
the rate coefficients for excitation were added to 
those calculated using the Free Electron model,
in the scattering-length approximation,
of \citet{1985JPhB...18L.167K,kaulakys1986free,1991JPhB...24L.127K}.
The Free Electron model was also used to estimate
the rate of ionisation by hydrogen impact.
The collisional data described above
do not consider fine structure; thus, the rate coefficients
within LS terms were set to very high values
while the rate coefficients between LS terms
were redistributed to the LSJ structure of the model atom
so as to approximately conserves the total rates
\citep[e.g.][]{2025A&A...696A.210C}.
Test 1D non-LTE calculations 
using a model atom with rate coefficients perturbed
by a factor of ten showed that
the statistical equilibrium is overall most sensitive
to the excitation by neutral hydrogen impact,
followed by excitation by electron impact. 
Moreover, the \ion{S}{I} $1045\,\nm$ triplet, which shows
large departures from LTE, is mainly
sensitive to the collisions that couple its own lower and upper levels.

To reduce the computational cost, a 
$66$ level reduced model atom was constructed.
Fine structure LSJ levels were collapsed
into LS terms for levels above $8.8\,\ev$,
and LS terms above $9.93\,\ev$
of the same multiplicity and within $\pm0.05\,\ev$ were merged
into so-called super levels. 
There are seven super levels in total,
with the highest one just $0.15\,\ev$ below 
the first ionisation limit.
The affected transitions were collapsed into super transitions
following the approaches described in \citet{2024ARA&A..62..475L}.
Test 1D non-LTE calculations
showed that the departure coefficients 
of the comprehensive and reduced model atoms
closely follow each other,
such that the abundance corrections for both the disc-centre
intensity and disc-integrated flux agree between the two 
models to $0.001\,\dex$ in the worst case.
For the production runs,
the non-LTE iterations were performed on the reduced model
to determine departure coefficients for each LSJ level,
LS term, or super level;
and the final emergent spectra were calculated after applying
these departure coefficients onto the LTE populations of
the corresponding LSJ levels or LS terms
in the comprehensive model.

\subsection{Diagnostic \ion{S}{I} lines and oscillator strengths}
\label{methodlggf}

\begin{table*}
\begin{center}
\caption{Parameters for the diagnostic \ion{S}{I} lines.}
\label{tab:linelist}
\begin{tabular}{l l l l l l | r r r r}
\hline
\hline
\noalign{\smallskip}
\multirow{2}{*}{$\lambda_{\mathrm{air}} / \nm$} & 
\multirow{2}{*}{Lower} & 
\multirow{2}{*}{Upper} & 
\multirow{2}{*}{$E_{\mathrm{low}} / \ev$} & 
\multirow{2}{*}{$\upsigma / a_{0}^2$} & 
\multirow{2}{*}{$\upalpha$} & 
\multicolumn{4}{c}{$\lggf$} \\ 
 & 
 & 
 & 
 & 
 & 
 & 
Exp. & \bsr{} & \civ{} & \grasp{} \\ 
\noalign{\smallskip}
\hline
\hline
\noalign{\smallskip}
$    469.54$
 & 
$\mathrm{               3p^{3}(^{4}S^{o})4s\, ^{5}S^{o}_{2}
}$ & 
$\mathrm{                   3p^{3}(^{4}S^{o})5p\, ^{5}P_{2}
}$ & 
$    6.5245$
 & 
$ 1280$
 & 
$0.267$
 & 
 
 & 
$     -1.87$
 & 
$     -1.75$
 & 
$     -1.83$
 \\ 
$    675.68$
 & 
$\mathrm{                   3p^{3}(^{4}S^{o})4p\, ^{5}P_{3}
}$ & 
$\mathrm{               3p^{3}(^{4}S^{o})5d\, ^{5}D^{o}_{2}
}$ & 
$    7.8699$
 & 
$ 2833$
 & 
$0.247$
 & 
 
 & 
$     -1.78$
 & 
 
 & 
$     -1.86$
 \\ 
$    675.70$
 & 
$\mathrm{                   3p^{3}(^{4}S^{o})4p\, ^{5}P_{3}
}$ & 
$\mathrm{               3p^{3}(^{4}S^{o})5d\, ^{5}D^{o}_{3}
}$ & 
$    7.8699$
 & 
$ 2832$
 & 
$0.248$
 & 
 
 & 
$     -0.94$
 & 
 
 & 
$     -1.01$
 \\ 
$    675.72$
 & 
$\mathrm{                   3p^{3}(^{4}S^{o})4p\, ^{5}P_{3}
}$ & 
$\mathrm{               3p^{3}(^{4}S^{o})5d\, ^{5}D^{o}_{4}
}$ & 
$    7.8699$
 & 
$ 2832$
 & 
$0.248$
 & 
 
 & 
$     -0.35$
 & 
 
 & 
$     -0.43$
 \\ 
$    867.02$
 & 
$\mathrm{                   3p^{3}(^{4}S^{o})4p\, ^{5}P_{1}
}$ & 
$\mathrm{               3p^{3}(^{4}S^{o})4d\, ^{5}D^{o}_{0}
}$ & 
$    7.8663$
 & 
$ 1417$
 & 
$0.302$
 & 
 
 & 
$     -0.91$
 & 
$     -0.83$
 & 
$     -0.92$
 \\ 
$    869.46$
 & 
$\mathrm{                   3p^{3}(^{4}S^{o})4p\, ^{5}P_{3}
}$ & 
$\mathrm{               3p^{3}(^{4}S^{o})4d\, ^{5}D^{o}_{4}
}$ & 
$    7.8699$
 & 
$ 1415$
 & 
$0.302$
 & 
 
 & 
$+      0.05$
 & 
$+      0.19$
 & 
$+      0.04$
 \\ 
$   1045.54$
 & 
$\mathrm{               3p^{3}(^{4}S^{o})4s\, ^{3}S^{o}_{1}
}$ & 
$\mathrm{                   3p^{3}(^{4}S^{o})4p\, ^{3}P_{2}
}$ & 
$    6.8601$
 & 
$  625$
 & 
$0.228$
 & 
$+      0.25$
 & 
$+      0.26$
 & 
$+      0.27$
 & 
$+      0.26$
 \\ 
$   1045.68$
 & 
$\mathrm{               3p^{3}(^{4}S^{o})4s\, ^{3}S^{o}_{1}
}$ & 
$\mathrm{                   3p^{3}(^{4}S^{o})4p\, ^{3}P_{0}
}$ & 
$    6.8601$
 & 
$  625$
 & 
$0.228$
 & 
$     -0.45$
 & 
$     -0.44$
 & 
$     -0.43$
 & 
$     -0.44$
 \\ 
$   1045.94$
 & 
$\mathrm{               3p^{3}(^{4}S^{o})4s\, ^{3}S^{o}_{1}
}$ & 
$\mathrm{                   3p^{3}(^{4}S^{o})4p\, ^{3}P_{1}
}$ & 
$    6.8601$
 & 
$  625$
 & 
$0.228$
 & 
$+      0.03$
 & 
$+      0.04$
 & 
$+      0.05$
 & 
$+      0.04$
 \\ 
\noalign{\smallskip}
\hline
\hline
\end{tabular}
\tablefoot{Hydrogen collisional broadening parameters based on ABO theory with
$\upsigma$ the cross-section at reference velocity $10^{4}\,\mathrm{m\,s^{-1}}$
and $\upalpha$ the exponent such that the cross-section goes as $v^{-\upalpha}$.
Experimental oscillator strengths (Exp.) from \citet{1997PhyS...56..459Z}, and
theoretical oscillator strengths (\bsr{}, \civ{}, \grasp{}) in the Babushkin
(length) gauge from \citet{2006JPhB...39.2861Z}, \citet{2008ADNDT..94..561D},
and \citet{li_submitted} respectively.}
\end{center}
\end{table*}

The focus of this work is on seven \ion{S}{I} lines,
a subset of the eight lines studied in
\citet{2015A&A...573A..25S}. 
The \ion{S}{I} $469.41\,\nm$ that was included in
their study is omitted here, because of difficulties
in accurately placing the continuum 
corresponding to a large uncertainty in equivalent width.
We highlight the seven \ion{S}{I} lines in \fig{fig:atom},
and illustrate their formation regions
in \fig{fig:atmos}.
We also list the key line parameters in \tab{tab:linelist}.
As the \ion{S}{I} $675\,\nm$ triplet is modelled
with fine structure, there are nine lines shown in the table.

Four sets of oscillator strengths are considered in this analysis
and are listed in different columns of \tab{tab:linelist}.
First are the experimental data
of \citet{1997PhyS...56..459Z}, which are only
available for the \ion{S}{I} $1045\,\nm$ triplet.  
These are based on laboratory lifetime measurements
using laser spectroscopy.
The reported uncertainties are only of the order $\pm0.01\,\dex$.

The other three data sets are 
based on mid-scale or large-scale theoretical atomic structure
calculations.
These are, in chronological order: the
$B$-Spline $R$-matrix (\bsr{}) calculations
of \citet{2006JPhB...39.2861Z};
the configuration interaction
calculations of \citet{2008ADNDT..94..561D} with the \civ{} code;
and the multiconfiguration Dirac-Hartree-Fock
(MCDHF) calculations of \citet{li_submitted}
with the \grasp{} code
\citep{2019CoPhC.237..184F,2023Atoms..11...68J},
following an approach that is similar 
to that described in \citet{2021MNRAS.502.3780L,
2023ApJS..265...26L,2023A&A...674A..54L}.
As discussed by  \citet{2008ADNDT..94..561D}, their
\civ{} calculations are about one order of magnitude larger
than the \bsr{} ones; 
while the \grasp{} ones are yet another order
of magnitude larger than the \civ{} ones, as estimated by the number
of employed configuration state functions (CSFs). 
The \civ{} calculations and the \grasp{} calculations both
employ fine-tuning of the Hamiltonian matrix
to match experimental energies
\citep{1996PhST...65..104H,2023Atoms..11...70L}.
The results from the Babushkin (relativistic) or 
length (non-relativistic) gauge are used
in this work.
The internal differences with the Coulomb gauge
(corresponding to the velocity gauge in the non-relativistic
limit) are typically
small for these lines (for the \grasp{} calculations the maximum difference
is $0.014\,\dex$).

The accuracy of the theoretical 
oscillator strengths (and other theoretical transition data)
depends on how accurately the wave functions of the
upper and lower state of the transition are described. If the
percentage purity of a state is high, it is easier to describe them accurately.
Conversely, if the mixing is strong, it is necessary to consider sufficient
configuration interaction to describe them accurately.
Consequently, if the upper or lower states, or both,
of the diagnostic lines are strongly mixed,
it can lead to strong cancellation
effects, which can hamper the accuracy of the computed transition data
when the configuration interaction is not sufficiently taken into account
\citep[e.g.][]{2023A&A...674A..54L}. 
This renders some diagnostic lines more reliable than others when
relying on theoretical oscillator strengths
\citep[e.g.][]{2021MNRAS.502.3780L}.

For the \ion{S}{I} $1045\,\nm$ triplet, 
the \grasp{} calculations \citep{li_submitted} indicate that
the lower and upper states have high percentage purity
and that the cancellation effects are not severe.
\tab{tab:linelist} shows that all three theoretical 
data sets agree with each other and with the experimental
results of \citet{1997PhyS...56..459Z} to better than $0.02\,\dex$. 
Consequently, consideration of these
experimental data alone do not strongly support
one theoretical data set over the other.

In contrast, the other lines
appear to have much more uncertain oscillator strengths.
\tab{tab:linelist} shows that
the \bsr{} and \grasp{} sets are systematically smaller than
the \civ{} set by around $0.1\,\dex$,
after averaging over the \ion{S}{I} 
$469.54\,\nm$, $867.02\,\nm$, and $869.46\,\nm$ lines.
In the \grasp{} calculations
these transitions display more severe cancellation effects 
than the \ion{S}{I} $1045\,\nm$ triplet,
as indicated cancellation factors that are closer to zero
\citep{1981tass.book.....C}.
The \ion{S}{I} $675\,\nm$ triplet is particularly uncertain,
as the \bsr{} and \grasp{} values differ by more
than $0.07\,\dex$,
and unfortunately no value is provided for it
in the \civ{} set. The upper level of this feature,
$\mathrm{3p^{3}(^{4}S^{o})5d\,^{5}D^{o}}$,
experiences significant mixing with other $\mathrm{3p^{3}}n\mathrm{d}$ levels
and the cancellation factor of the three components are all
less than $0.1$ which indicates a high degree of cancellation
\citep[e.g.][]{2023A&A...674A..54L,2024EPJD...78...36K}.
The uncertainties in the theoretical oscillator strengths as suggested
by the differences between the three data sets
play a significant role when trying to validate
the abundance analysis on the basis of systematic differences
between different groups of lines (\sect{resultsline})
and thus to infer the most reliable solar sulphur abundance
(\sect{resultsfinal}).

The emergent spectra for the diagnostic \ion{S}{I} lines were
only calculated with a single set of oscillator strengths, namely those given in the
comprehensive model atom.  For the 
\ion{S}{I} $1045\,\nm$ triplet these are the precise
experimental values of \citet{1997PhyS...56..459Z},
while for the other lines these oscillator strengths come
from the \bsr{} calculations of
\citet{2006JPhB...39.2861Z} in the Babushkin gauge.
As such, for a given line,
the solar sulphur abundance inferred from the different atomic
data sets was evaluated by adjusting the abundances inferred from
comprehensive model atom for the difference in
oscillator strength, via
$\Delta \lgeps{S} = -\Delta \lggf$.

\subsection{Observational data}
\label{methoddata}

\begin{table*}
\begin{center}
\caption{Equivalent widths and abundance corrections determined in this work.}
\label{tab:abcor}
\begin{tabular}{l | r r | r r | r r | r r | r r}
\hline
\hline
\noalign{\smallskip}
\multirow{3}{*}{$\lambda_{\mathrm{air}} / \nm$} & 
\multirow{3}{*}{$W_{\mathrm{I}} / \mathrm{pm}$} &
\multirow{3}{*}{$W_{\mathrm{F}} / \mathrm{pm}$} &
\multicolumn{8}{c}{Abundance correction} \\ 
 & 
 & 
 & 
\multicolumn{2}{c|}{$\mathrm{1N-1L}$} & \multicolumn{2}{c|}{$\mathrm{3N-3L}$} & \multicolumn{2}{c|}{$\mathrm{3N-1N}$} & \multicolumn{2}{c}{$\mathrm{3N-1L}$} \\ 
 & 
 & 
 & 
\multicolumn{1}{c}{I} & \multicolumn{1}{c|}{F} & \multicolumn{1}{c}{I} & \multicolumn{1}{c|}{F} & \multicolumn{1}{c}{I} & \multicolumn{1}{c|}{F} & \multicolumn{1}{c}{I} & \multicolumn{1}{c}{F} \\
\noalign{\smallskip}
\hline
\hline
\noalign{\smallskip}
$    469.54$
 & 
$      0.87$
 & 
$      0.74$
 & 
$     -0.01$
 & 
$     -0.02$
 & 
$     -0.02$
 & 
$     -0.03$
 & 
$     -0.01$
 & 
$     -0.01$
 & 
$     -0.02$
 & 
$     -0.04$
 \\ 
$    675.71$
 & 
$      2.33$
 & 
$      1.81$
 & 
$     -0.01$
 & 
$     -0.02$
 & 
$     -0.02$
 & 
$     -0.03$
 & 
$      0.01$
 & 
$     -0.01$
 & 
$      0.00$
 & 
$     -0.02$
 \\ 
$    867.02$
 & 
$      0.60$
 & 
$      0.47$
 & 
$     -0.02$
 & 
$     -0.02$
 & 
$     -0.02$
 & 
$     -0.04$
 & 
$      0.00$
 & 
$     -0.02$
 & 
$     -0.01$
 & 
$     -0.04$
 \\ 
$    869.46$
 & 
$      3.40$
 & 
$      2.73$
 & 
$     -0.02$
 & 
$     -0.04$
 & 
$     -0.04$
 & 
$     -0.06$
 & 
$      0.03$
 & 
$     -0.01$
 & 
$      0.01$
 & 
$     -0.04$
 \\ 
$   1045.54$
 & 
$     13.40$
 & 
$     11.30$
 & 
$     -0.11$
 & 
$     -0.19$
 & 
$     -0.14$
 & 
$     -0.23$
 & 
$      0.07$
 & 
$      0.02$
 & 
$     -0.05$
 & 
$     -0.17$
 \\ 
$   1045.68$
 & 
$      6.20$
 & 
$      5.42$
 & 
$     -0.07$
 & 
$     -0.12$
 & 
$     -0.09$
 & 
$     -0.15$
 & 
$      0.05$
 & 
$      0.01$
 & 
$     -0.02$
 & 
$     -0.12$
 \\ 
$   1045.94$
 & 
$     10.60$
 & 
$      9.35$
 & 
$     -0.10$
 & 
$     -0.17$
 & 
$     -0.13$
 & 
$     -0.21$
 & 
$      0.06$
 & 
$      0.02$
 & 
$     -0.04$
 & 
$     -0.16$
 \\ 
\noalign{\smallskip}
\hline
\hline
\end{tabular}
\tablefoot{The \ion{S}{I} $675.7\,\nm$ triplet is treated here as a single blended feature.}
\end{center}
\end{table*}

The analysis of the solar sulphur abundance
in \sect{resultsline} and \sect{resultsfinal}
is primarily based on
equivalent widths measured in the
Li\`ege \citep{1973apds.book.....D} and Hamburg
\citep{1984SoPh...90..205N} disc-centre intensity atlases,
$W_{\mathrm{I}}$.
These data were collected at high altitude observatories
making them less susceptible to telluric contamination,
and with instruments of extremely high resolving power
making the impact of blends easier to judge and mitigate.
Small differences were found between the equivalent widths
measured from the two atlases,
of the order a few percent.  As these differences
did not appear to be systematic, it is valid to adopt the mean
value as done by 
\citet{2015A&A...573A..25S}.

Lines in the disc-centre intensity spectrum form deeper in the solar atmosphere
than lines in the the disc-integrated flux spectrum,
as seen by the contribution functions to the line depressions
illustrated in \fig{fig:atmos}.
They are therefore usually less susceptible to 
3D and non-LTE modelling uncertainties.
The lines in disc-centre intensity are 
also not affected by rotational broadening
and so blends are easier to identify
and isolate. Thus, the advocated solar sulphur
abundance we give in \sect{resultsfinal}
is based on the disc-centre intensity.
Nevertheless, the disc-integrated flux is also 
considered in \sect{resultsratio} as the comparison
of spectra forming in different parts of the solar
atmosphere can help validate the 3D non-LTE modelling.
For the disc-integrated flux, the equivalent widths 
$W_{\mathrm{F}}$ were measured
in the IAG atlas \citep{2016A&A...587A..65R}.

The adopted $W_{\mathrm{I}}$ and $W_{\mathrm{F}}$ are listed in \tab{tab:abcor}.
It should be noted that although
the \ion{S}{I} $675\,\nm$ triplet is modelled
with fine structure (\tab{tab:linelist}), the components significantly
overlap in the solar spectrum, and a single
equivalent width is listed corresponding to the entire feature.
For all seven lines,
the $W_{\mathrm{I}}$ were taken from \citet{2015A&A...573A..25S},
after finding them to be consistent with our values to
better than $5\%$. 
Concerning the flux measurements,
our $W_{\mathrm{F}}$ are in good agreement
with those of \citet{2005PASJ...57..751T}
for the four lines in common,
namely the \ion{S}{I} $869.46\,\nm$ and the
\ion{S}{I} $1045\,\nm$ triplet;
the largest disagreement is for the \ion{S}{I} $1045.94\,\nm$ line,
where our value is $6\%$ larger than that of 
\citet{2005PASJ...57..751T}.

\section{Results}
\label{results}

\begin{figure*}
    \begin{center}
        \includegraphics[scale=0.325]{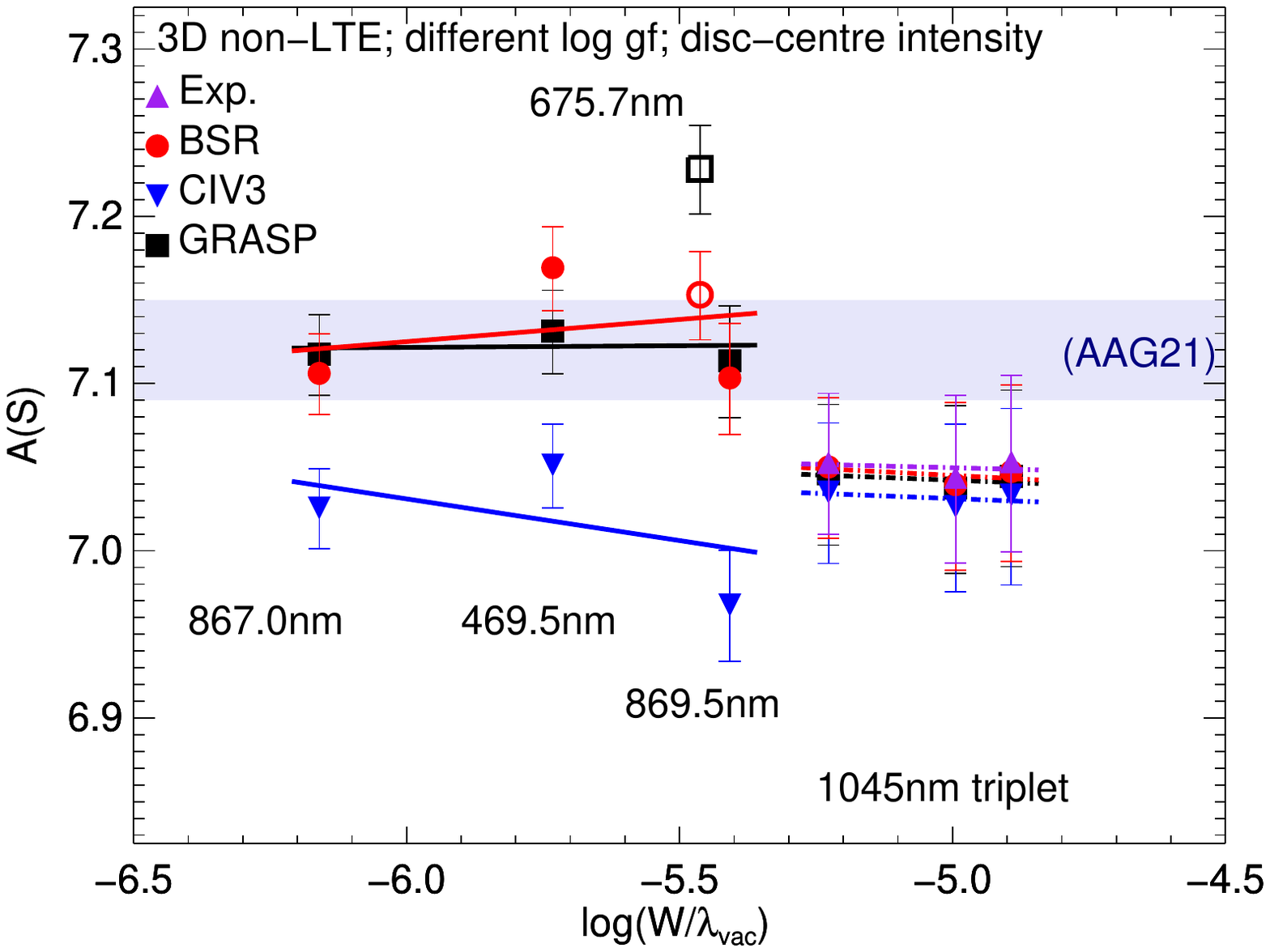}\includegraphics[scale=0.325]{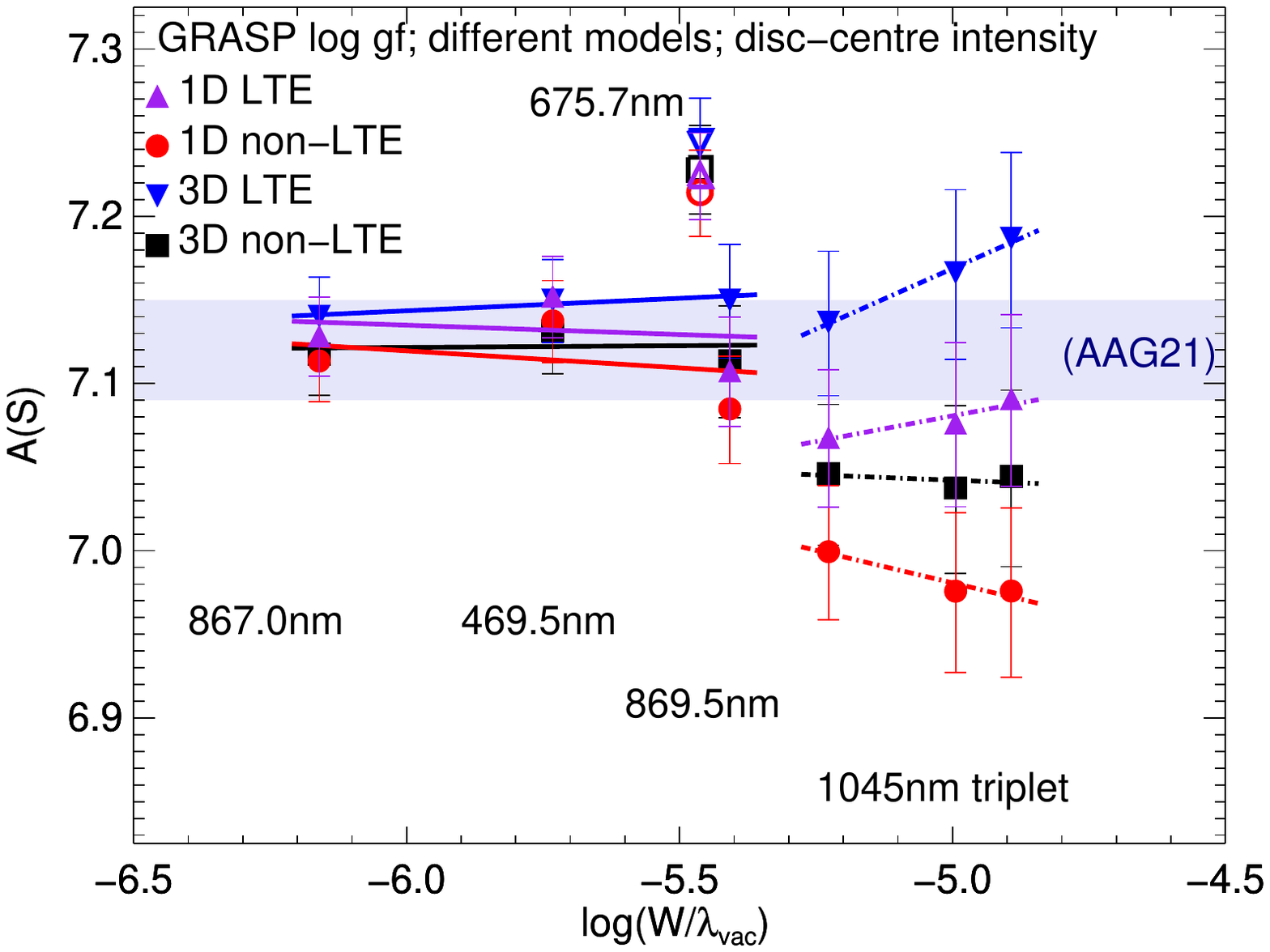}
        \includegraphics[scale=0.325]{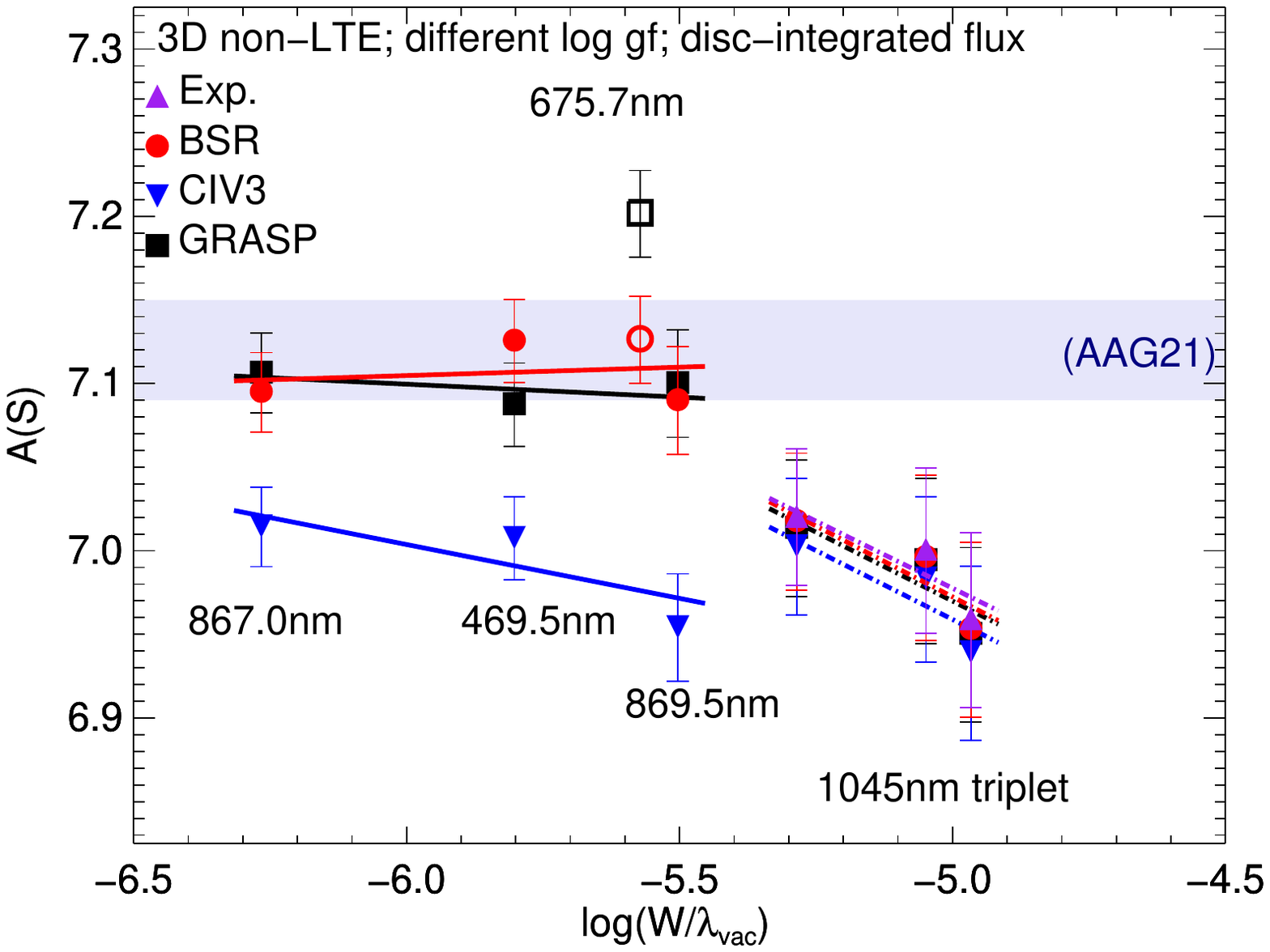}\includegraphics[scale=0.325]{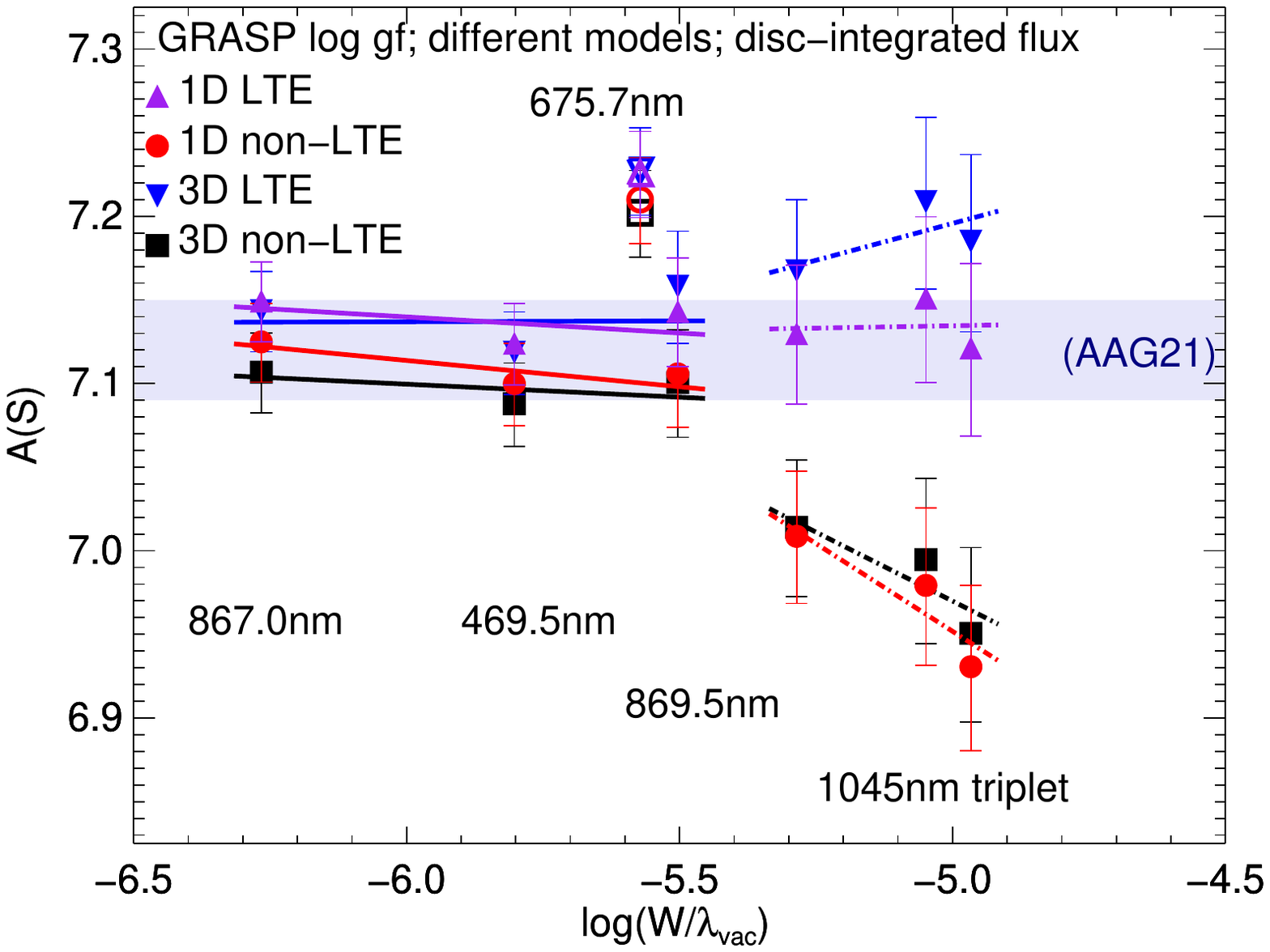}
        \caption{Abundances inferred from different sulphur lines
        as a function of logarithmic reduced equivalent width.
        Upper row shows results from the disc-centre intensity,
        and lower row shows results from the disc-integrated
        flux.
        Left panel shows results for different
        sets of oscillator strengths based on the 3D non-LTE model.
        Right panel shows results for different models
        based on \grasp{} oscillator strengths.
        Black squares are the same in the left and right panels
        of a given row.
        Error bars reflect the estimated $\pm5\%$ uncertainty
        on the measured equivalent widths.
        Weighted linear regressions to different line groups
        shown, with the \ion{S}{I} $675\,\nm$ triplet 
        (not present in the \civ{} set; open symbols)
        given zero weight in the fits. 
        The shaded rectangle shows the solar sulphur abundance
        given in \citet{2021A&A...653A.141A} of
        $\lgeps{S}=7.12\pm0.03$.}
        \label{fig:abundances}
    \end{center}
\end{figure*}

\begin{figure*}
    \begin{center}
        \includegraphics[scale=0.2225]{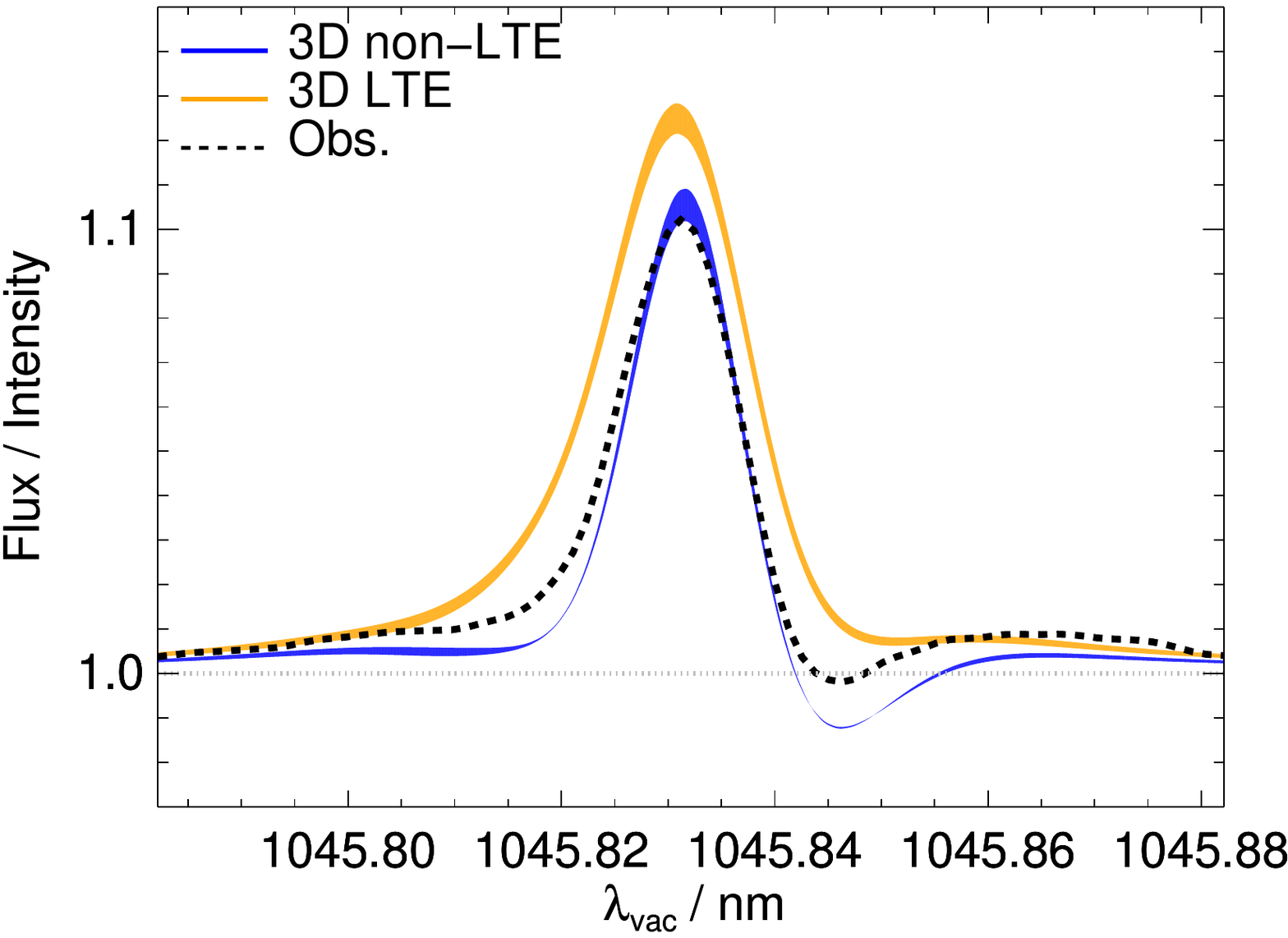}\includegraphics[scale=0.2225]{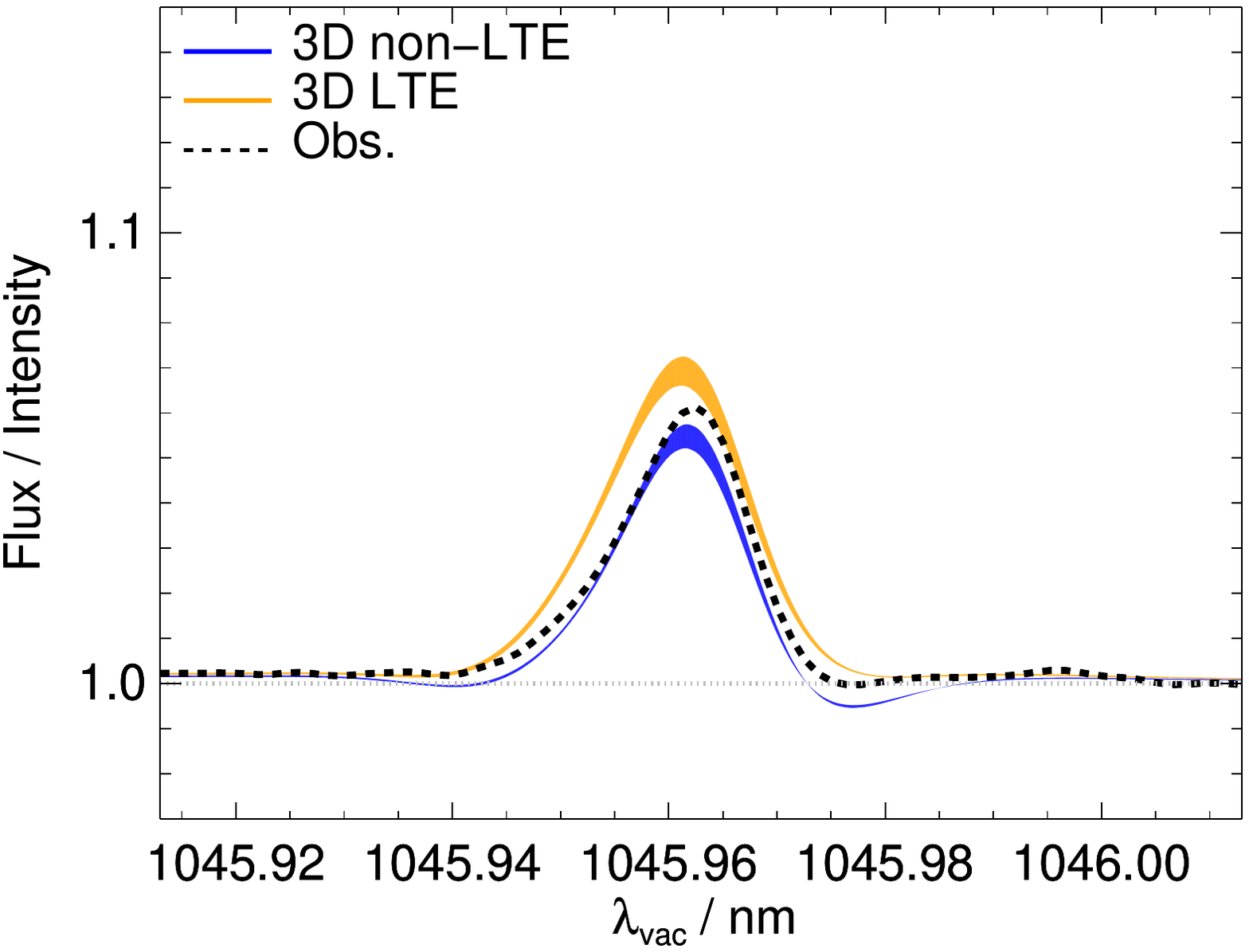}\includegraphics[scale=0.2225]{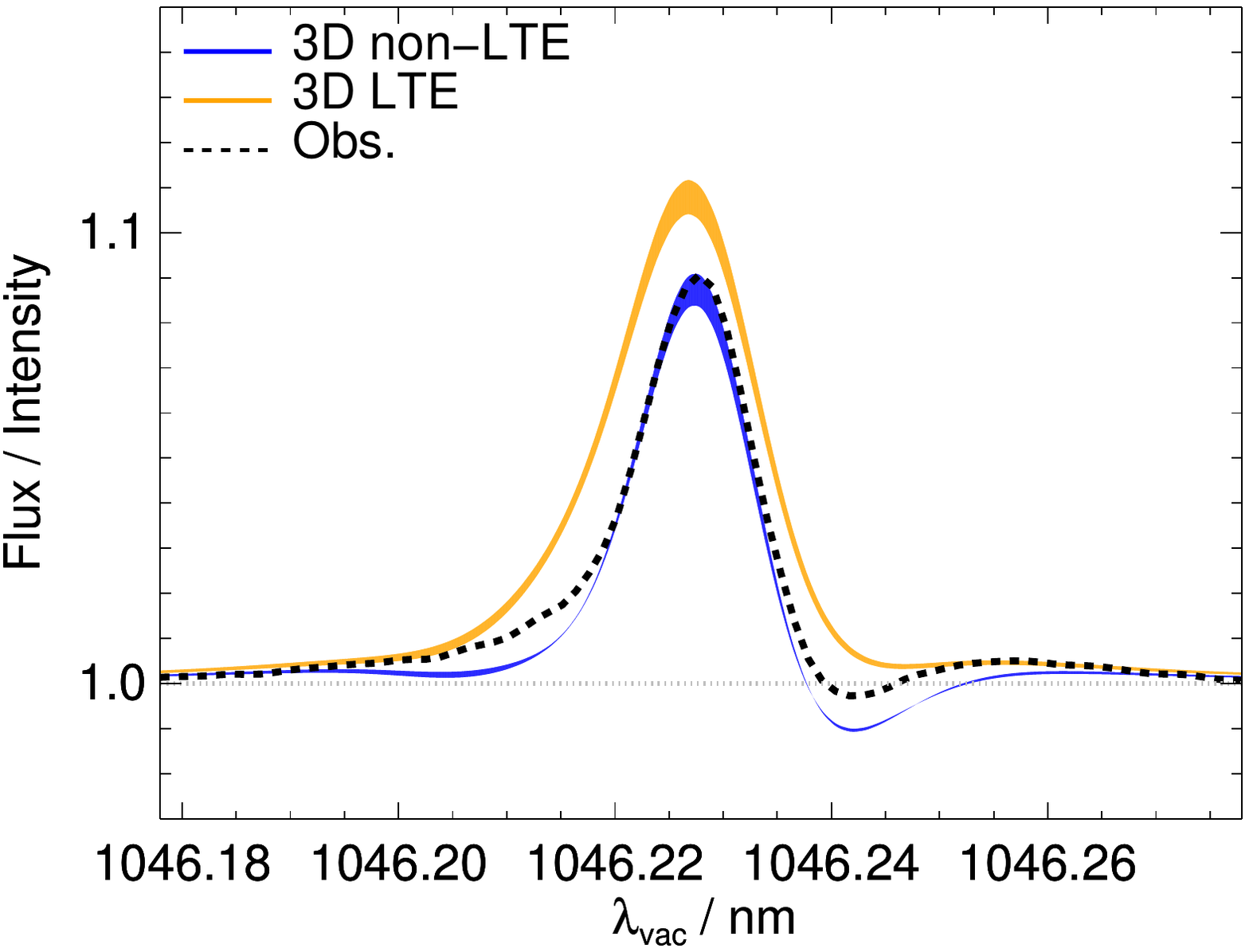}
        \caption{Disc-integrated flux to disc-centre intensity ratios
        for the \ion{S}{I} $1045\,\nm$ triplet.  
        Observations from the Hamburg atlas.
        The thickness of the plotted lines
        corresponds to the $\pm1\sigma$ uncertainty 
        in the 3D non-LTE and 3D LTE abundances 
        illustrated in \fig{fig:abundances}, that 
        results from a $\pm5\%$ uncertainty in the
        measured equivalent widths.
        Instrumental broadening
        of $R=3.5\times10^{5}$ included for the
        synthetic intensities and fluxes,
        and rotational broadening of $\vsini=2\,\kms$
        included for the synthetic flux.}
        \label{fig:ratio}
    \end{center}
\end{figure*}

\subsection{Abundance corrections}
\label{resultsabcor}

We provide the derived abundance corrections
for the seven \ion{S}{I} lines
in the disc-centre intensity 
and in the disc-integrated flux
in \tab{tab:abcor}.
These were derived by determining the
synthetic equivalent widths by direct integration and 
finding the abundances that matched the observed equivalent widths 
($W_{\mathrm{I}}$ or $W_{\mathrm{F}}$)
in 1D LTE ($\mathrm{1L}$),
1D non-LTE ($\mathrm{1N}$),
3D LTE ($\mathrm{3L}$),
and 3D non-LTE ($\mathrm{3N}$).
The columns of the table give four different abundance corrections
determined as the difference in these fitted abundances:
$\mathrm{1N-1L}$, 
$\mathrm{3N-3L}$ (the non-LTE effect),
$\mathrm{3N-1N}$ (the 3D effect),
and $\mathrm{3N-1L}$.

The abundance corrections in disc-centre intensity
are fairly similar for the weak
\ion{S}{I} $469.54\,\nm$, $675.71\,\nm$, and $867.02\,\nm$ lines.
These form at similar heights in the atmosphere
(\fig{fig:atmos}) and are in the same spin system
(\fig{fig:atom}).
The 3D and non-LTE effects for these lines
are usually small, in the range $-0.02\,\dex$ to
$+0.01\,\dex$.
The \ion{S}{I} $869.46\,\nm$ line is slightly stronger in terms
of reduced equivalent width. It forms higher up in the atmosphere
(\fig{fig:atmos}) and has slightly more severe abundance corrections;
the non-LTE effect is $\mathrm{3N-3L}=-0.04\,\dex$
and the 3D effect is $\mathrm{3N-1N}=+0.03\,\dex$.
As expected from previous work, 
the lines which have the most severe abundance corrections 
are the \ion{S}{I} $1045\,\nm$ triplet lines, with the blue-most
component having a non-LTE effect of $-0.14\,\dex$
and a 3D effect of $+0.07\,\dex$ in the disc-centre intensity.
The effects going in different directions which means
that the $\mathrm{3N-1L}$ difference is not extremely severe,
only $-0.05\,\dex$. It is worth noting that
the components suffer different 3D and non-LTE effects because
they form at different depths
(as discussed in e.g.~\citealt{2024A&A...689A..36K}).

The 3D and non-LTE
effects tend to be even more severe in the disc-integrated
flux than in the disc-centre intensity.
The \ion{S}{I} $1045\,\nm$ triplet presents a notable exception,
where the 3D effect is much reduced in the disc-integrated flux.
It should be noted that
this result is sensitive to the microturbulence
fudge parameter adopted in the 1D models.
Although a value of $\vmic=1\,\kms$ is typical
for the disc-centre intensity and is adopted here,
a larger value may in fact be more appropriate for the disc-integrated
flux \citep[e.g.][]{2022SoPh..297....4T}.
Increasing the microturbulence 
increases the strength of the 1D non-LTE 
synthesis, which makes $\mathrm{3N-1N}$ more positive.

It is interesting to now consider which
diagnostics are likely to be the more reliable abundance diagnostics.
As we discussed in \sect{methodlggf},
the oscillator strengths for the
\ion{S}{I} $1045\,\nm$ triplet are well-constrained
with precise laboratory measurements of precisions
$\pm0.01\,\dex$
as well as the three different theoretical data sets 
in agreement with these measurements
to $0.02\,\dex$.
However, this advantage is offset by
their higher sensitivity to the
modelling approach.
In contrast, the weaker lines show much less severe abundance corrections,
with the $\mathrm{3N-1L}$ difference always in the 
range $-0.02$ to $+0.01\,\dex$ in the disc-centre intensity.
In other words, the sensitivity to the models
for these lines is relatively weak.
However, as we discussed in \sect{methodlggf},
the oscillator strengths for these lines
are poorly constrained.  There do not exist
precise laboratory measurements, and the theoretical \civ{} set
is systematically offset from the 
\bsr{} and \grasp{} sets by around $0.1\,\dex$.
The two groups of lines thus have distinct
advantages and disadvantages, which should
be taken into account when deriving a final estimate
of the solar sulphur abundance (\sect{resultsfinal}).

\subsection{Line-by-line abundances from equivalent widths in
disk-centre intensity}
\label{resultsline}

We show the abundances inferred from different lines
in the disc-centre intensity
in the two upper panels of \fig{fig:abundances}.  
As we discussed in \sect{resultsabcor}, the
\ion{S}{I} $1045\,\nm$
triplet lines have well-constrained oscillator strengths (small scatter in the
upper left panel), 
but a relatively large model dependence (large scatter in the upper right
panel).  Conversely, the other lines have poorly-constrained oscillator
strengths, but a relatively small model dependence.

The 3D non-LTE model gives consistent sulphur abundances
from the three components of the \ion{S}{I} $1045\,\nm$ triplet,
whereas as shown in the upper right panel of \fig{fig:abundances},
the other modelling approaches result in steeper trends
in the inferred sulphur abundance with reduced equivalent width.
Furthermore, if the 3D non-LTE model is assumed to be accurate,
then the upper left panel suggests that the \civ{} data
are more reliable out of the three theoretical
data sets, as this set results in consistent results (on average)
for the other lines together with the \ion{S}{I} $1045\,\nm$ triplet.

On the other hand, the better overall
agreement between the \bsr{} and \grasp{} results may indicate that these
data sets are more reliable than the \civ{} set of values.
Moreover, comparisons with experimental lifetimes
also suggest that the \bsr{} and \grasp{} sets
are slightly more reliable overall compared
to \civ{} (see \citealt{li_submitted}), although these comparisons
do not include the lower or upper levels of the weak
lines considered here.
If these data sets are more reliable on average, then
the upper right panel of \fig{fig:abundances} suggests that
the \ion{S}{I} $1045\,\nm$ triplet is best modelled in 3D LTE,
with 3D non-LTE in systematic error by around $0.1\,\dex$.
This would be surprising given that 3D non-LTE 
has consistently been demonstrated to outperform all other
models, and especially 3D LTE, in various contexts
\citep[e.g.][]{2024ARA&A..62..475L}. Furthermore,
the analysis we present in \sect{resultsratio} also favours
3D non-LTE over 3D LTE.
Nevertheless, independent 3D non-LTE modelling
of \ion{S}{I} lines would be welcome to help
verify our results.

\subsection{Comparison with results from the disc-integrated flux} 
\label{resultsratio}

The spectral lines in the disc-integrated flux form in higher layers
than the spectral lines in the disc-centre intensity.
Consequently, the disc-integrated flux
can also be analysed as a consistency check, one that is sensitive to the
atmospheric stratification (3D effects)
as well as for the non-LTE modelling given that
departures from LTE tend to grow with height in the solar atmosphere.
It should be noted, however, that uncertainties
in the equivalent width measurements,
in particular placing the continuum in
crowded parts of the spectrum, can make the results
of this consistency test less clear.

Overall, the disc-integrated flux results in the lower 
left panel of \fig{fig:abundances}
shows a similar pattern to the 
disc-centre intensity results in the
upper left panel of that figure.
Namely, with 3D non-LTE modelling,
the \ion{S}{I} $1045\,\nm$ triplet
indicates a low solar sulphur
abundance for all four different sets of oscillator 
strengths. Meanwhile, the weaker lines give
(on average) consistent results with
the triplet when adopting the \civ{} 
set of oscillator strengths.

The mean result for the 
\ion{S}{I} $1045\,\nm$ triplet 
in the disc-integrated flux is
$6.99\,\dex$ with a standard deviation of $0.03\,\dex$,
while in the disc-centre
intensity it is $7.05\,\dex$ with negligible scatter, based on the
experimental oscillator strengths 
of \citet{1997PhyS...56..459Z}.
For the other lines,
based on the \civ{} oscillator strengths
(and excluding the \ion{S}{I} $675\,\nm$ triplet for
which there are no data in the \civ{} set)
the average result in the disc-integrated flux is
$6.99\,\dex$ with a standard deviation of $0.03\,\dex$, 
while in the disc-centre
intensity it is $7.01\,\dex$ with a standard deviation of 
$0.04\,\dex$.

Overall there is good agreement between
the results from the disc-centre intensity
and the disc-integrated flux.
However, there is a hint of a systematic bias
towards lower inferred abundances in the latter
case.  In particular, for the
\ion{S}{I} $1045\,\nm$ triplet, the difference of
$0.06\,\dex$ is somewhat larger
than the line-to-line scatter.
This could indicate a residual 
shortcoming in the 3D non-LTE models, such that
the 3D non-LTE effects are slightly overestimated.
Such an error would affect the synthesis of the disc-integrated flux
to a greater degree than the disc-centre intensity,
as the latter forms deeper in the atmosphere.

The ratio of the disc-integrated fluxes and disc-centre intensities
can be used as an additional test of the 
\ion{S}{I} $1045\,\nm$ triplet.
This follows \citet{2019A&A...624A.111A},
who considered centre-to-limb ratios
for carbon lines using the centre-to-limb atlas of \citet{2015A&A...573A..74S}.
Here, the observational data were drawn from
the normalised Hamburg atlas
\citep{1984SoPh...90..205N} for the flux and for the intensity.
The additional free parameters 
microturbulence and macroturbulence limit the
usefulness of such a comparison for
the 1D non-LTE and 1D LTE models, and so we limit
our attention to the 3D non-LTE and 3D LTE models here.
The disc-integrated fluxes were 
broadened for stellar rotation following \citet{1990A&A...228..203D} and 
adopting $\vsini=2.0\,\kms$.
Instrumental broadening was also applied to
the emergent disc-centre intensities and disc-integrated fluxes,
assuming $R=3.5\times10^{5}$
\citep{1999SoPh..184..421N} adopting a sinc$^{2}$ kernel.
No other additional broadening was applied.
The abundances were set to those inferred 
from the analysis of the equivalent widths in disc-centre intensity,
namely 
$7.05$ for the blue and middle components,
and $7.04$ for the red component
(\tab{tab:abund}).

We show the flux to intensity ratios in \fig{fig:ratio}. 
In comparison to observations,
the 3D non-LTE model outperforms the 3D LTE model for
each of the components of the \ion{S}{I} $1045\,\nm$ triplet. 
It should be kept in mind that this test is carried
out without the inclusion of any blending species.
Still, the 3D non-LTE model
shows some deviations in the wings, particularly
for the strongest component for which the core also lies
slightly above the observed data.
As the  \ion{S}{I} $1045\,\nm$ triplet is sensitive
to stellar magnetic fields 
\citep[e.g.][]{2024A&A...689A..36K}, 
these deviations in the wings of the ratio may be caused
by the Zeeman effect,
which is neglected in the present model, and thus
the theoretical intensity profiles are too narrow. 
Interestingly, this is opposite to what is found by some other
groups, where the line profiles
from 3D \cobold{} models are sometimes too broad
\citep[e.g.][]{2022A&A...668A..48D}.
The effect of some missing Zeeman broadening
on the present abundance analysis is mitigated by
working with equivalent widths rather than spectral line fits.
Moreover previous studies of other magnetically
sensitive lines such as the \ion{O}{I} $777\,\nm$ triplet
suggest that the effects of small-scale magnetic fields
in the solar atmosphere are less than $0.01\,\dex$ in abundances
inferred from equivalent widths
\citep[e.g.][]{2016A&A...586A.145S}.

Nevertheless, in future work, it may
be worthwhile to explore 3D non-LTE spectrum synthesis 
of the \ion{S}{I} $1045\,\nm$ triplet
using 3D radiative-magnetohydrodynamics models of the solar photosphere.
It may also be interesting to investigate
the impact of uncertainties in the line broadening
on the statistical equilibrium, although this
likely corresponds to a small error overall,
given that the 1D non-LTE versus 1D LTE abundance corrections
for the \ion{S}{I} $1045\,\nm$ triplet 
changes by less than $0.01\,\dex$ effect when increasing
the microturbulence from $1\,\kms$ to $2\,\kms$.

\subsection{Line-averaged solar sulphur abundance}
\label{resultsfinal}

\begin{table*}
\begin{center}
\caption{Adopted oscillator strengths and recommend line-by-line and line-averaged solar sulphur abundances for different models, in the disc-centre intensity and disc-integrated flux.}
\label{tab:abund}
\begin{tabular}{l l | r r | r r | r r | r r}
\hline
\hline
\noalign{\smallskip}
\multirow{3}{*}{$\lambda_{\mathrm{air}} / \nm$} & 
\multirow{3}{*}{$\lggf$} & 
\multicolumn{8}{c}{Inferred abundance} \\ 
& & 
\multicolumn{2}{c|}{$\mathrm{1L}$} & \multicolumn{2}{c|}{$\mathrm{1N}$} & \multicolumn{2}{c|}{$\mathrm{3L}$} & \multicolumn{2}{c}{$\mathrm{3N}$} \\ 
& & 
\multicolumn{1}{c}{I} & \multicolumn{1}{c|}{F} & \multicolumn{1}{c}{I} & \multicolumn{1}{c|}{F} & \multicolumn{1}{c}{I} & \multicolumn{1}{c|}{F} & \multicolumn{1}{c}{I} & \multicolumn{1}{c}{F} \\
\noalign{\smallskip}
\hline
\hline
\noalign{\smallskip}
$    469.54$
 & 
$     -1.79$
 & 
$      7.11$
 & 
$      7.08$
 & 
$      7.10$
 & 
$      7.06$
 & 
$      7.11$
 & 
$      7.08$
 & 
$      7.09$
 & 
$      7.05$
 \\ 
$    675.71$
 & 
$     -0.24$
 & 
$      7.15$
 & 
$      7.15$
 & 
$      7.14$
 & 
$      7.14$
 & 
$      7.17$
 & 
$      7.16$
 & 
$      7.16$
 & 
$      7.13$
 \\ 
$    867.02$
 & 
$     -0.87$
 & 
$      7.08$
 & 
$      7.10$
 & 
$      7.07$
 & 
$      7.08$
 & 
$      7.09$
 & 
$      7.10$
 & 
$      7.07$
 & 
$      7.06$
 \\ 
$    869.46$
 & 
$+      0.11$
 & 
$      7.03$
 & 
$      7.07$
 & 
$      7.01$
 & 
$      7.03$
 & 
$      7.08$
 & 
$      7.09$
 & 
$      7.04$
 & 
$      7.03$
 \\ 
$   1045.54$
 & 
$+      0.25$
 & 
$      7.10$
 & 
$      7.13$
 & 
$      6.98$
 & 
$      6.94$
 & 
$      7.20$
 & 
$      7.19$
 & 
$      7.05$
 & 
$      6.96$
 \\ 
$   1045.68$
 & 
$     -0.45$
 & 
$      7.07$
 & 
$      7.14$
 & 
$      7.01$
 & 
$      7.02$
 & 
$      7.14$
 & 
$      7.17$
 & 
$      7.05$
 & 
$      7.02$
 \\ 
$   1045.94$
 & 
$+      0.03$
 & 
$      7.08$
 & 
$      7.16$
 & 
$      6.98$
 & 
$      6.99$
 & 
$      7.17$
 & 
$      7.22$
 & 
$      7.04$
 & 
$      7.00$
 \\ 
\noalign{\smallskip}
\hline
\noalign{\smallskip}
\multicolumn{2}{c|}{Line-average} & 
$      7.08$
 & 
$      7.10$
 & 
$      7.03$
 & 
$      7.04$
 & 
$      7.12$
 & 
$      7.11$
 & 
$      7.06$
 & 
$      7.03$
 \\ 
 & & & & & & & & 
$\pm      0.04$
 & 
$\pm      0.05$ \\
\noalign{\smallskip}
\hline
\hline
\end{tabular}
\tablefoot{See \sect{resultsfinal} for the explanation of the adopted $\lggf$ values and the line averaging procedure. The \ion{S}{I} $675.7\,\nm$ triplet, given zero weight in the line average, is shown for completeness; its quoted $\lggf$ is the logarithm of the sum of $gf$ values of the three components. }
\end{center}
\end{table*}

We present line-by-line abundance results
in \tab{tab:abund}, based on the disc-centre intensity
and disc-integrated flux.
To derive a line-averaged result,
the lines were separated into two groups
that suffer from different dominant systematic
uncertainties as discussed in \sect{resultsline}.
The line-averaged result was found by 
averaging in two groups separately.

The first group contains
the three \ion{S}{I} $1045\,\nm$ lines.
These have a high sensitivity to
the models (3D and non-LTE effects),
but they also have precise oscillator strengths based on
laboratory measurements \citep{1997PhyS...56..459Z}.
The uncertainty in the line-averaged abundance was estimated as
half the mean non-LTE effect
$\mathrm{3N-3L}$ added in quadrature with half the mean
3D effect $\mathrm{3N-1N}$.
The first group gives $\lgeps{S}=7.050\pm0.067$
in the 
disc-centre intensity.

The second group contains the other, weaker lines.
These have only a small sensitivity
to the models, but a large uncertainty
in their oscillator strengths as suggested
by the $0.1\,\dex$ systematic shift between the 
\civ{} set 
\citep{2008ADNDT..94..561D},
compared to the \grasp{} set \citep{li_submitted} 
and the \bsr{} set
\citep{2006JPhB...39.2861Z}.
It is unclear which set is most reliable
(\sect{resultsline}).
To take this systematic uncertainty into account,
the oscillator strengths
from the \civ{} and \grasp{} sets
were averaged together.
Using the \bsr{} set instead of the \grasp{} set
here modifies the final result by less than $0.01\,\dex$.
As the \ion{S}{I} $675\,\nm$ triplet is not included
in the \civ{} data, and the 
\grasp{} and \bsr{} values have a large difference of $0.07\,\dex$
that may reflect uncertainties caused by significant cancellation effects
due to mixing of the 
$\mathrm{3p^{3}(^{4}S^{o})5d\,^{5}D^{o}}$ state
(\sect{methodlggf}), this feature is deemed too questionable and
is given zero weight in the line-averaged result. 
Nevertheless, for completeness, we list
the abundances inferred from this feature in \tab{tab:abund};
this is based on the average of the \grasp{} and 
\bsr{} oscillator strengths,
reduced by $0.053\,\dex$ which is
half of the systematic difference between the two data sets
averaged over the three weak lines.
The uncertainty in the line-averaged abundance
was then estimated as half of the average systematic difference
($0.053\,\dex$)
added in quadrature with the standard error,
which reflects the random uncertainties
in the oscillator strengths and
equivalent width measurements.
The second group gives $\lgeps{S}=7.068\pm0.055$
in the 
disc-centre intensity.

The bottom of \tab{tab:abund} displays the 
resulting weighted mean and uncertainty 
after averaging the two groups
using weights proportional to $\sigma^{-2}$.
The 3D non-LTE result from the disc-centre intensity
is $\lgeps{S}=7.06\pm0.04$.
The same weights were used to generate
line-averaged results for the three other models.
Repeating this exercise for the disc-integrated flux
gives  $\lgeps{S}=6.994\pm0.10$ for the first group
and $\lgeps{S}=7.045\pm0.054$ for the second group,
corresponding to a weighted mean and uncertainty
of $\lgeps{S}=7.03\pm0.05$ in 3D non-LTE. 
It is reassuring that this result is consistent with that from
the disc-centre intensity to within $1\sigma$.
This is expected because we have
attempted to fold in both random measurement 
errors as well as systematic modelling uncertainties.

\section{Discussion}
\label{discussion}

\begin{figure}
    \begin{center}
        \includegraphics[scale=0.325]{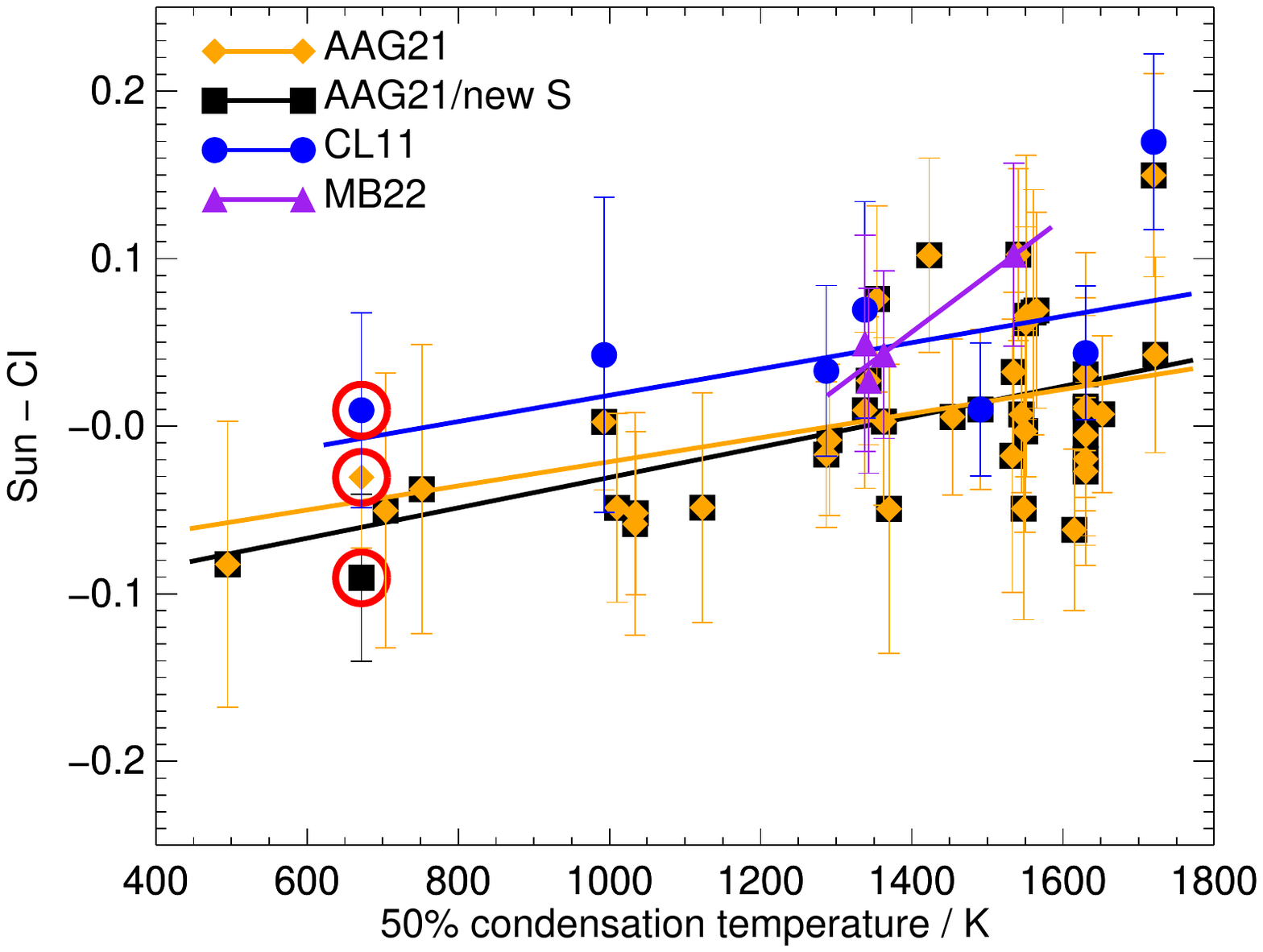}
        \caption{Photospheric versus CI chondrite
        abundance differences as a function of
        $50\%$ condensiation temperature from
        \citet{2019AmMin.104..844W}.
        Photospheric data AAG21, CL11, and MB22 are
        from \citet{2021A&A...653A.141A},
        \citet{2011SoPh..268..255C},
        and \citet{2022A&A...661A.140M}
        respectively, the latter restricted
        to their derivations based on a
        horizontally- and temporally-averaged 3D model.
        CI chondrite data is from 
        \citet{2021SSRv..217...44L},
        converted to the solar scale
        with $\lgeps{Si}=7.51$ from \citet{2021A&A...653A.141A}.
        Only elements with combined
        uncertainties less severe than $\pm0.1\,\dex$ are included,
        and the reference element silicon is omitted
        from those data sets where it was available.
        Sulphur, with $\tcond=672\,\kelvin$, has been circled,
        and the new solar sulphur abundance found 
        in this work is used to update AAG21,
        forming the data set AAG21/new S.
        Weighted linear regressions are overplotted.}
        \label{fig:tcond}
    \end{center}
\end{figure}

\subsection{Comparison with the solar sulphur abundance of
\citet{2015A&A...573A..25S} and \citet{2021A&A...653A.141A}}
\label{discussionscott}

The line-by-line 3D LTE results presented in \tab{tab:abund} agree with those
of \citet{2015A&A...573A..25S} to within $0.02\,\dex$ on average,
after correcting for differences in the adopted oscillator strengths
and including the \ion{S}{I} $469.41\,\nm$ line.
The 1D LTE results agree to $0.01\,\dex$ on average.
Some of this residual discrepancy may
originate from differences between the 3D model atmosphere, 
the one used here
being computed with the solar composition of 
\citet{2009ARA&A..47..481A}, while that of 
\citet{2015A&A...573A..25S} is based on an older
model computed with the more metal-poor composition
of \citet{2005ASPC..336...25A}.
Some of the other differences may reflect
updates in the spectrum synthesis with \scate{},
in particular to the atomic data and partition functions
as described in \citet{2021A&A...656A.113A}.

Concerning the non-LTE modelling,
it is interesting to note that the $\mathrm{1N-1L}$ corrections
presented in \tab{tab:abcor}
for different lines are all within $0.02\,\dex$ of
the $\mathrm{1N-1L}$ corrections
in disc-centre intensity given in
\citet{2015A&A...573A..25S}
based on the model of \citet{2005PASJ...57..751T}.
It should be noted that 
\citet{2015A&A...573A..25S} calibrated their 
non-LTE model so as to reduce the line-to-line scatter,
by scaling all of their inelastic hydrogen
collisions (computed with the Drawin recipe) by a factor of $0.4$.
In this work the non-LTE model takes a purely first principles approach
that includes a more physically-motivated treatment of 
the inelastic hydrogen collisions,
and therefore no calibrations were applied.

Our advocated solar sulphur abundance
is different by $7.06-7.12=-0.06\,\dex$ compared to that 
presented in \citet{2015A&A...573A..25S}
and \citet{2021A&A...653A.141A}.
The difference is somewhat larger
than the $1\sigma$ uncertainty
of $\pm0.04\,\dex$ arrived at here,
as well as the combined $1\sigma$ uncertainty of
$\sqrt{0.04^2+0.03^2}=0.05$
(even though these two uncertainties
are not independent).
Part of the difference is due to using
a consistent 3D non-LTE approach in this work.
Using an inconsistent 3D LTE + 1D non-LTE approach
(3D LTE abundances with 1D non-LTE
abundance corrections) as done by \citet{2015A&A...573A..25S}
raises our result by $0.02\,\dex$, 
putting it within $1\sigma$ of their result.

\subsection{Comparison with the solar sulphur abundance of
\citet{2007A&A...470..699C} and \citet{2011SoPh..268..255C}}
\label{discussioncaffau}

Another commonly-used compilation of the solar
chemical composition is that of
\citet{2011SoPh..268..255C}, which lists
$\lgeps{S}=7.16\pm0.05$.
The same value appears to be adopted in the compilation of
\citet{2025SSRv..221...23L}
but with a much larger stipulated uncertainty of 
$\pm0.11\,\dex$ in their Table 2.

The result of \citet{2011SoPh..268..255C} is based on the weak
[\ion{S}{I}] $1082\,\nm$ line
\citep{2007A&A...467L..11C}, which
we discuss in \sect{discussionmorelines},
as well as
the \ion{S}{I} $675\,\nm$ triplet,
the \ion{S}{I} $869.39\,\nm$ and $869.46\,\nm$ lines,
and the \ion{S}{I} $1045\,\nm$ triplet
\citep{2007A&A...470..699C}.
For the latter, they carried out a 3D LTE analysis 
of the disc-integrated flux. They adopted
1D non-LTE versus 1D LTE abundance corrections
from \citet{2005PASJ...57..751T}. 
This model employs the Drawin recipe for the
inelastic hydrogen collisions with neutral
hydrogen, without any downscaling.
As such the abundance corrections for
the disc-integrated flux used by
\citet{2007A&A...470..699C} are less severe
than the abundance corrections for the disc-centre
intensity of \citet{2015A&A...573A..25S}, 
which adopted a scaling of $0.4$ as we 
discussed in \sect{discussionscott}.

Our advocated solar sulphur abundance
(\sect{resultsfinal})
is different by $7.06-7.16=-0.10\,\dex$ compared to that 
presented in \citet{2011SoPh..268..255C}.
Adding the uncertainties of 
$\pm0.04\,\dex$
and $\pm0.05\,\dex$ in quadrature,
the difference amounts to more than $1.5\sigma$.

One source of the differences may be
the non-LTE modelling for the \ion{S}{I} $1045\,\nm$ triplet.
Our 1D non-LTE versus 1D LTE abundance corrections 
for the disc-integrated flux
are around $0.10\,\dex$ more negative than the
values of $-0.09\,\dex$, $-0.05\,\dex$, and $-0.07\,\dex$,
that they adopted from \citet{2005PASJ...57..751T}.

Another source may be the measurement of the solar
spectrum. 
It is convenient to consider the logarithmic ratios of
equivalent widths
in the disc-integrated flux spectrum,
dividing those given in Table 2 of
\citet{2007A&A...470..699C} by
those in our \tab{tab:abcor}.
For the five lines in common, these ratios are
between $0.02\,\dex$ and $0.10\,\dex$;
our values being systematically smaller.
Likewise, the equivalent widths given in their Table 2
tend to be larger than those given in Table 4 of \citet{2005PASJ...57..751T}.
In fact, the equivalent widths for the disc-integrated
flux reported in \citet{2007A&A...470..699C}
are very close to our values for the disc-centre intensity 
in our \tab{tab:abcor}.
When repeating our analysis of the disc-integrated flux but
adopting their equivalent widths,
our 3D LTE results agree with theirs to better than
$0.01\,\dex$ after averaging the lines in common.
Correcting for differences in oscillator strengths
slightly worsens the agreement, to $0.02\,\dex$.

\subsection{Other diagnostic \ion{S}{I} 
lines of astrophysical interest.}
\label{discussionmorelines}

It is worth briefly discussing the 
[\ion{S}{I}] $1082.1\,\nm$ line
(\sect{discussionmorelinesforbidden})
and the \ion{S}{I} $920\,\nm$ triplet
(\sect{discussionmorelinesallowed}), even though
they were ultimately not folded into our advocated solar abundance.
The [\ion{S}{I}] $1082.1\,\nm$ line
is sometimes used to study Galactic chemical evolution
via sulphur abundances in giant stars
\citep[e.g.][]{2006A&A...455L..13R,
2011A&A...530A.144J,2013A&A...559A.115M},
and the \ion{S}{I} $920\,\nm$ triplet
has been used for both giants
\citep[e.g.][]{2023A&A...671A.137L}
and dwarfs 
\citep[e.g.][]{2007A&A...469..319N}
usually at lower metallicity.

\subsubsection{The [\ion{S}{I}] $1082.1\,\nm$ line}
\label{discussionmorelinesforbidden}

As we mentioned in \sect{discussioncaffau},
the [\ion{S}{I}] $1082.1\,\nm$ line
was studied in \citet{2007A&A...467L..11C}
and the results of which were folded
into the recommended value of the solar
sulphur abundance given in \citet{2011SoPh..268..255C}.
They adopted $\lggf=-8.617$ from an old version of the NIST database;
this value is close to 
published theoretical values of $-8.615$ \citep{1983MNRAS.202..981M}
and $-8.601$ \citep{1986PhyS...34..116B}.
Their mean equivalent width is
$W_{\mathrm{I}}=0.218\,\mathrm{pm}$
With these values,
our 3D LTE abundance is $\lgeps{S}=7.15$.
Our 1D abundance is only around $0.02\,\dex$ smaller,
and non-LTE effects were confirmed to be negligible for this line
at least in the solar spectrum.
These results are in good quantitative agreement with those
reported in Table 2 of \citet{2007A&A...467L..11C}.

However, the equivalent width of the line is difficult to precisely
determine in the solar spectrum.  It is strongly blended on its red side, and
sits on the outer wing of a strong feature.  Our estimate is
$W_{\mathrm{I}}\approx0.28\,\mathrm{pm}$, corresponding
to a $0.11\,\dex$ larger inferred abundance. Concerning the oscillator strength,
the NIST database currently states $\lggf=-8.73$ with an accuracy rating of
``C'' (around $\pm0.1\,\dex$).  This is based on the theoretical calculation of
\citet{2006ADNDT..92..607F} but corrected by around $-0.03\,\dex$ for the
experimental wavelength.  Section 6.6 of \citet{2015A&A...573A..25S} states a
value of $\lggf=-8.774$ after Froese-Fischer (priv.~comm.).  Adopting this
latter value would raise the inferred abundance a further $0.16\,\dex$.

To shed 
light on this, we carried out a separate \grasp{} calculation for the
[\ion{S}{I}] $1082.1\,\nm$ M1 and E2 transitions.
The model adopted a multi-reference set encompassing 
$\mathrm{3s^{2}3p^{4}}$, $\mathrm{3s3p^{4}3d}$,
$\mathrm{3p^{6}}$, and $\mathrm{3s^{2}3p^{2}3d^{2}}$ to produce the CSF
expansion to account for the important interactions.
The calculations were carried out in a layer-by-layer scheme,
expanding the size of the active set of correlation orbitals up to
$n = 7$ and $l = 5$ in the $J=2$ symmetry block,
corresponding to a maximum of $263\,363$ CSFs. 
A systematic investigation was performed on the convergence of the oscillator
strength with respect to the choice of multi-reference, electron correlation
effects, and the size of the active set of correlation orbitals. From this
analysis, we recommend a combined M1+E2 
oscillator strength corresponding to $\lggf=-8.625$.

With $W_{\mathrm{I}}\approx0.28\,\mathrm{pm}$ and
our \grasp{} $\lggf=-8.625$,
we arrive at $\lgeps{S}=7.27$, which is difficult
to reconcile with the results from the 
allowed \ion{S}{I} lines regardless of which atomic data
set are adopted for them.
Despite our new oscillator strength, we
arrive at the same conclusion as
in Section 6.6 of \citet{2015A&A...573A..25S}, that 
the weak [\ion{S}{I}] $1082.1\,\nm$ line
may be significantly blended at least in the Sun.

\subsubsection{The \ion{S}{I} $920\,\nm$ triplet}
\label{discussionmorelinesallowed}

The \ion{S}{I} $920\,\nm$ triplet
($921.29\,\nm$, $922.81\,\nm$, and $923.75\,\nm$)
is known to be affected by telluric water absorption
when viewed using ground-based telescopes.
In the IAG disc-integrated flux atlas
this renders them unusable.
Fortunately, the lines appear much cleaner
in the Li\`ege disc-centre intensity atlas.
The middle component has an uncertain 
continuum placement;  
from the blue and red components we could estimate equivalent widths
of $W_{\mathrm{I}}\approx16.44\,\mathrm{pm}$
and $10.80\,\mathrm{pm}$, respectively.
In the \grasp{} set the three components
have $\lggf=0.395$, $0.248$, and $0.026$ 
\citep{li_submitted}, in close agreement with the \bsr{} values
via the NIST database; the lines are not present
in the \civ{} set.
The inferred 3D non-LTE abundance becomes $\lgeps{S}\approx6.99$,
thereby also favouring a low solar sulphur abundance.

It must be stressed that the \ion{S}{I} $920\,\nm$ triplet 
is saturated in the solar disc-centre intensity spectrum,
with logarithmic reduced equivalent widths 
of $-4.93$ for the weakest component,
and $-4.75$ for the strongest component.
They are also extremely sensitive to the models,
with $\mathrm{3N-3L}\approx-0.18$ in the disc-centre intensity
and reaching almost $0.3\,\dex$ in the disc-integrated flux.
The overall 3D non-LTE effect is slightly less severe,
with $\mathrm{3N-1L}\approx-0.07$ in the disc-centre intensity
and around $-0.2\,\dex$ in the disc-integrated flux.
For these reasons this feature is considered an unreliable
solar abundance diagnostic.

\subsection{Comparison with the 1D non-LTE modelling
of \citet{2024ARep...68.1159K}}
\label{discussionkorotin}

The non-LTE modelling of the $1045\,\nm$ triplet
plays an important role in shaping the conclusions of the current study.
Thus, it is a useful consistency check to
compare our model with that of
an independent study.
Although there are no other 3D non-LTE results in the literature,
recently \citet{2024ARep...68.1159K}
presented a detailed 1D non-LTE study of neutral and ionised
sulphur in AFGK-type stars.
They used \atlas{} model atmospheres
\citep[e.g.][]{2003IAUS..210P.A20C}
and the \multi{} 1D non-LTE code
\citep[e.g.][]{1986UppOR..33.....C}, and an
independent approach to constructing the model atom. 
The complexity of the model atom in that work
is similar to that used here, and
key data sources which the statistical
equilibrium is strongly sensitive to,
including excitation by neutral
hydrogen impact 
\citep{2020ApJ...893...59B}
and electron impact
\citep{2011AIPC.1344..179S},
are the same in their model as in ours.
Their model does not take into account
fine structure, which can have a small impact on 
the statistical equilibrium and thus
the abundance corrections \citep[e.g.][]{2024A&A...690A.128A}.

To make a quantitative comparison 
with \citet{2024ARep...68.1159K},
1D non-LTE calculations were carried out
on the \marcs{} solar model atmosphere \citep{2008A&A...486..951G}.
To better mimic their approach,
$\vmic=2.0\,\kms$ was adopted.
Furthermore, the equivalent widths
were determined by integrating
across the $1045\,\nm$ triplet as a whole as done in
the grid of \citet{2024ARep...68.1159K},
although the $\mathrm{1N-1L}$
values listed in \tab{tab:abcor} reveal
that the different components may suffer
rather different non-LTE effects.

The resulting 1D non-LTE versus 1D LTE abundance
correction for the $1045\,\nm$ triplet in
the disc-integrated flux,
at an abundance of $\lgeps{S}=7.14$,
comes to $-0.17\,\dex$.
Importantly, if repeating this same test
but calculating the statistical equilibrium
using a version of the 
reduced model atom in which all fine structure
is collapsed (but still carrying out the final
spectrum synthesis on the same comprehensive
model atom described
in \sect{methodatom}), the abundance
correction becomes slightly less severe,
at $-0.15\,\dex$.

The latter result is in excellent agreement
with the grid of \citet{2024ARep...68.1159K}, 
which is calculated without fine structure.
Interpolating their grid
to the solar effective temperature and surface gravity, and adopting 
$\mathrm{[Fe/H]}=0$ and $\mathrm{[S/Fe]}=0$,
gives a 1D non-LTE versus 1D LTE abundance
correction of $-0.14\,\dex$. 
This is just $0.01\,\dex$ away from the result obtained here
when using the reduced model atom with no fine structure, 
which is closer to the approach
of \citet{2024ARep...68.1159K}.
This agreement makes us more confident about our,
non-LTE modelling and, consequently, about
the overall conclusions of this study.

In lieu of full 3D non-LTE synthetic spectra or abundance corrections,
the 1D non-LTE approach often gives more reliable results than 1D LTE
\citep{2024ARA&A..62..475L}.
Thus, we provide an extended grid of 1D non-LTE departure coefficients
for sulphur levels on standard \marcs{} model atmospheres.
These 1D non-LTE calculations employed the reduced model atom
(including fine structure), and
follow those presented previously in the literature
for other elements \citep[e.g.][]{2024A&A...687A...5M,2025A&A...696A.210C}
with background scattering treated as in
\citet{2022A&A...668A..68A}.
They span the stellar parameter space described 
in \citet{2020A&A...642A..62A}, but here
with $-1.5\leq\mathrm{[S/Fe]}\leq+1.5$ using steps of $0.5\,\dex$.
The data are provided\footnote{\url{https://doi.org/10.5281/zenodo.17064337}}
as text files as well as in formats
that readily work with the codes \sme{} 
\citep{2017A&A...597A..16P}
and \pysme{} \citep{2023A&A...671A.171W},
with the usual caveat that the user must take care to ensure the labels
of the sulphur levels in their linelist
match those within the grid of data.

\subsection{Systematic differences between the solar photosphere 
and CI chondrites}

Although they have long been assumed to be equal,
different authors have recently pointed out the possibility
of a systematic difference between the composition of CI chondrites 
and of the solar photosphere
\citep[e.g.][]{2010MNRAS.407..314G,
2018ApJS..238...11D,2024M&PS...59.3193J}.
All of these studies use spectroscopic results from
\citet{2009ARA&A..47..481A} or \citet{2021A&A...653A.141A},
but \citet{2018ApJS..238...11D} added astrophysical modelling
based on the compositions and ages of solar system objects while
\citet{2024M&PS...59.3193J}
included solar wind values from the Genesis Solar Wind Sample Return
(NASA Discovery 5), as well as a suite of other spectroscopic studies.
The deviation between the compositions of 
the solar photosphere and CI chondrites are still being debated,
the question being if solar spectroscopy is accurate
enough to resolve such differences
\citep[e.g.][]{2025SSRv..221...23L}.

In \fig{fig:tcond} we plot the difference between the elemental
composition of the solar photosphere and of CI chondrites
($\mathrm{Sun-CI}$) 
as a function of $50\%$ condensation temperature.
For the photosphere,
several different data sets are used,
including \citet{2021A&A...653A.141A} and
\citet{2011SoPh..268..255C}
which we discussed in the context of the sulphur
abundance in \sect{discussionscott}
and \sect{discussioncaffau} respectively and are based
on 3D model atmospheres.
A third data set is that of \citet{2022A&A...661A.140M},
drawing on their internally-derived results 
which are based on 1D model atmospheres
constructed from a horizontally- and temporally-averaged 3D model.
A fourth data set is that of \citet{2021A&A...653A.141A}, 
but with the sulphur abundance
modified from $\lgeps{S}=7.12\pm0.03$
to $\lgeps{S}=7.06\pm0.04$ as we advocated
in \sect{resultsfinal}.
For the CI chondrites, the data set
of \citet{2021SSRv..217...44L} is used, converted to the solar
scale using $\lgeps{Si}=7.51$, the 3D non-LTE estimate from 
\citet{2017MNRAS.464..264A} that is adopted in \citet{2021A&A...653A.141A}.
The reference element silicon is omitted
from those data sets where it was available.
Also, following \citet{2021A&A...653A.141A}, only elements
with combined uncertainties in the
photospheric and CI chondrite abundances (added in quadrature)
less severe than $\pm0.1\,\dex$ are 
included.

In \fig{fig:tcond} we show the
weighted linear regressions for each of these data sets.
By eye, all sets of photospheric data appear consistent with
a linear $\mathrm{Sun-CI}$ versus $50\%$ condensation temperature.
According to the weighted linear regression,
the gradient given by the data of \citet{2021A&A...653A.141A}
is $2.4\sigma$ away from a flat line,
and that of \citet{2011SoPh..268..255C} is $1.3\sigma$ away
from a flat line.
In contrast, although the line connecting them in 
\fig{fig:tcond} is sloped, 
the \citet{2022A&A...661A.140M} data 
are scant and clustered around just two values of
$50\%$ condensation temperature that are only $200\,\kelvin$ apart.
So, given their moderate uncertainties, this set is
consistent with having zero gradient at $1\sigma$.

When updating the photospheric 
sulphur abundance in \citet{2021A&A...653A.141A}
to that derived in this work (AAG21/new S in \fig{fig:tcond}),
the weighted linear regression becomes steeper.
The gradient is now $2.9\,\sigma$ away from a flat line,
compared to $2.4\,\sigma$ based on the unaltered 
\citet{2021A&A...653A.141A} data set.
This is because the sulphur abundance
helps to drive the error-weighted trend.
With $\tcond=672\,\kelvin$, sulphur is one of relatively few
moderately volatile elements, and its abundance has relatively
low uncertainty (with $\pm0.04\,\dex$ derived in this work).
The only other elements in \fig{fig:tcond}
with $\tcond<800\,\kelvin$ and combined (photospheric and CI chondrite)
uncertainties less severe than $\pm0.1\,\dex$
are lead ($\tcond=495\,\kelvin$), 
zinc ($\tcond=704\,\kelvin$), and 
rubidium ($\tcond=752\,\kelvin$).
In \citet{2021A&A...653A.141A} these elements
have more severe uncertainties than that of sulphur:
$\pm0.08\,\dex$, $\pm0.05\,\dex$, and $\pm0.08\,\dex$ respectively.
Furthermore, these are all based on 3D LTE + 1D non-LTE 
analyses  rather than consistent 3D non-LTE,
and may therefore be more susceptible to
systematic errors.

In all, the present study supports the case for systematic
differences between the solar photosphere and CI chondrites
that are correlated with $50\%$ condensation temperature.
However, as discussed by \citet{2024M&PS...59.3193J}, the uncertainties 
are large enough that rather different profiles
may fit the $\mathrm{Sun-CI}$ versus
$50\%$ condensation temperature data equally well.
Using a linear abundance scale (rather than logarithmic as used here),
their study quantified the quality of different profiles using
the reduced chi-squared statistic, or mean squared weighted deviation,
following \citet{1991CGIGS..86..275W}. They found that several different 
profiles are consistent with the \citet{2021A&A...653A.141A} data,
including, for example, a straight line as used here; 
as well as a step function
with a break at around $\tcond=1350\,\kelvin$. When the new sulphur abundance
is substituted into the set of \citet{2021A&A...653A.141A},
the conclusions of \citet{2024M&PS...59.3193J} do not change
significantly, but the results are more consistent 
with the most probable value for each profile 
as quantified by the reduced chi-squared statistic.
More precise and accurate solar abundance measurements,
particularly for other moderately volatile elements,
would help with constraining different profiles
and subsequently understanding the origins
of the $\mathrm{Sun-CI}$ differences.

\section{Conclusion}
\label{conclusion}

The solar sulphur abundance has been reappraised.
After considering different sets of theoretical
oscillator strengths as well as, 
for the first time, consistent 3D non-LTE modelling,
we argue that the abundances of 
$\lgeps{S}=7.12\pm0.03$ in the compilation of 
\citet{2021A&A...653A.141A} 
and $\lgeps{S}=7.16\pm0.05$ in the compilation of 
\citet{2011SoPh..268..255C} are overestimated.
We advocate instead $\lgeps{S}=7.06\pm0.04$.

To put this argument on firmer footing,
independent 3D non-LTE determinations
would be welcome.
Such efforts would clearly benefit from
precise laboratory measurements of the oscillator strengths
of the \ion{S}{I} $469\,\nm$, $675\,\nm$, $867\,\nm$,
and $869\,\nm$ triplets.
Furthermore, it would be worth to repeat this study in the future
using 3D non-LTE post-processing of 
3D radiative-magnetohydrodynamics models of the solar photosphere,
to confirm the results obtained from
the \ion{S}{I} $1045\,\nm$ triplet.

The lower solar sulphur abundance supports
the case for a systematic difference
between the composition of the solar photosphere
and of CI chondrites 
that is correlated with $50\%$ condensation temperature
\citep{2010MNRAS.407..314G,2021A&A...653A.141A,2024M&PS...59.3193J}.
This result has implications for our understanding of the 
early history of the solar system, but remains debated.
To confirm the existence of this correlation
and better constrain its profile,
consistent 3D non-LTE abundance
analyses of other moderately volatile
elements would be welcome,
in particular for lead, zinc, and rubidium.

\begin{acknowledgements}
    We
    thank the referee (Elisabetta Caffau) for constructive feedback that
    helped improve the manuscript.
    We also thank Sema Caliskan for valuable comments on the manuscript.
    AMA acknowledges support from the Swedish Research Council (VR 2020-03940)
    the Crafoord Foundation via the Royal Swedish Academy of Sciences (CR
    2024-0015), and the European Union’s Horizon Europe research and innovation
    programme under grant agreement No. 101079231 (EXOHOST).  WL acknowledges
    the support from the specialized research fund for State Key Laboratory of
    Solar Activity and Space Weather.  AMA and WL also acknowledge support from
    the 2024 Chinese Academy of Sciences (CAS) President’s International
    Fellowship Initiative (PIFI).  AJGJ acknowledges support from the NASA
    Laboratory Analysis of Returned Samples (LARS) Program (award number
    80NSSC22K0589).  This research was supported by computational resources
    provided by the Australian Government through the National Computational
    Infrastructure (NCI) under the National Computational Merit Allocation
    Scheme and the ANU Merit Allocation Scheme (project y89).  
\end{acknowledgements}

\bibliographystyle{aa_url} 
\bibliography{bibl.bib}
\end{document}